\documentclass{article}
\usepackage{natbib}
\usepackage{epubtk}
\usepackage{graphicx}
\usepackage{rotating}
\usepackage{amsmath}
\usepackage{color}

\showlistoftablesfalse

\begin{document}

\title{Solar Force-free Magnetic Fields}

\author{%
\epubtkAuthorData{Thomas Wiegelmann}{%
Max-Planck Institut f\"ur Sonnensystemforschung \\
Max-Planck-Strasse 2 \\
37191 Katlenburg-Lindau \\
Germany}{%
wiegelmann@mps.mpg.de}{%
http://www.mps.mpg.de/homes/wiegelmann/ }%
\and
\epubtkAuthorData{Takashi Sakurai}{%
Solar and Plasma Astrophysics Division\\
National Astronomical Observatory of Japan \\
Mitaka, Tokyo 181-8588 \\
Japan}{%
sakurai@solar.mtk.nao.ac.jp}{%
http://solarwww.mtk.nao.ac.jp/sakurai/en/}
}

\date{}
\maketitle

\begin{abstract}
The structure and dynamics of the solar corona is dominated by the magnetic
field. In most areas in the corona magnetic forces are so dominant that all
non-magnetic forces like plasma pressure gradient and gravity can be neglected
in the lowest order. This model assumption is called the force-free field
assumption, as the Lorentz force vanishes. This can be obtained by either
vanishing electric currents (leading to potential fields) or the currents
are co-aligned with the magnetic field lines. First we discuss a
mathematically simpler approach that the magnetic field and currents
are proportional with one global constant, the so-called linear force-free
field approximation. In the generic case, however, the relation between
magnetic fields and electric currents is nonlinear and analytic solutions
have been only found for special cases, like 1D or 2D configurations. For
constructing realistic nonlinear force-free coronal magnetic field models
in 3D, sophisticated numerical computations are required and boundary
conditions must be obtained from measurements of the magnetic field
vector in the solar photosphere. This approach is currently of large
interests, as accurate measurements of the photospheric field become
available from ground-based (for example SOLIS) and space-born (for
example Hinode and SDO) instruments.  If we can obtain
accurate force-free coronal magnetic field models we can
calculate the free magnetic energy in the corona, a quantity which is
important for the prediction of flares and coronal mass
ejections. Knowledge of the 3D structure of magnetic field lines also
help us to interpret other coronal observations, e.g., EUV-images of
the radiating coronal plasma.
\end{abstract}
\epubtkKeywords{solar corona, magnetic fields}


\newpage
\section{Introduction}
\label{section:introduction}

The magnetic activity of the Sun has a high impact on Earth. As illustrated
in Figure~\ref{soho1}, large coronal eruptions like flares and coronal mass
ejections can influence the Earth's magnetosphere where they trigger magnetic
storms and cause aurorae. These coronal eruptions have also harmful effects
like disturbances in communication systems, damages on satellites, power
cutoffs, and unshielded astronauts are in danger of life-threatening
radiation.%
\epubtkFootnote{For an animation of a coronal mass ejection (CME)
  causing a substorm and aurora, see
  http://sohowww.nascom.nasa.gov/gallery/Movies/recon/reconsm.mpg
  }
The origin of these eruptive phenomena in the solar corona is
related to the coronal magnetic field as magnetic forces dominate
over other forces (like pressure gradient and gravity) in the
corona. The magnetic field, created by the solar dynamo, couples the
solar interior with the Sun's surface and atmosphere. Reliable high
accuracy magnetic field measurements are only available in the
photosphere. These measurements, called vector magnetograms,
provide the magnetic field vector in the photosphere.

\epubtkImage{soho}{%
\begin{figure}[htb]
\centerline{\includegraphics[width=\textwidth]{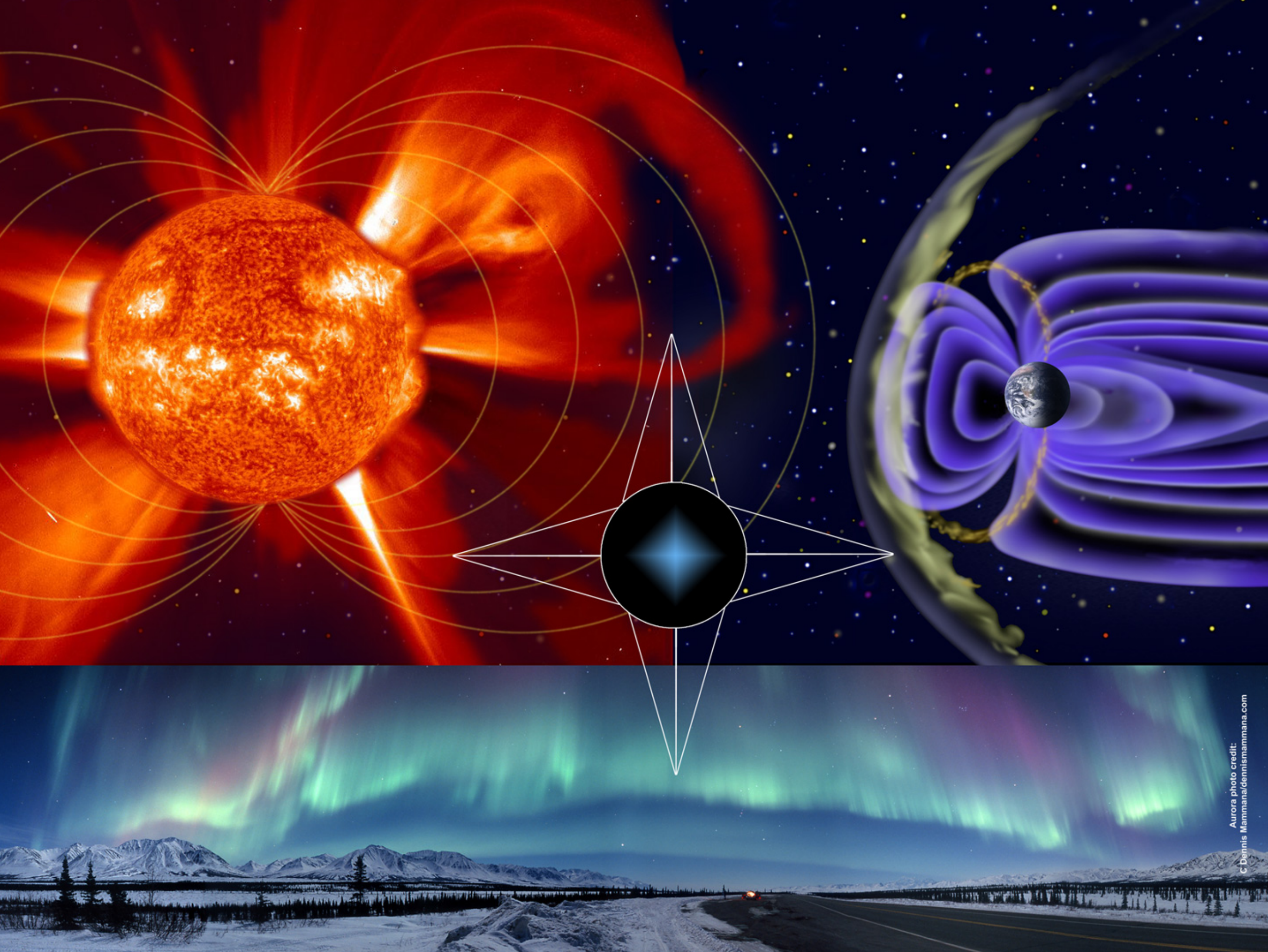}}
\caption{Magnetic forces play a key role in solar storms that can impact
Earth's magnetic shield (magnetosphere) and create colorful aurora. [Source:
http://sohowww.nascom.nasa.gov/gallery/images/magnetic\_clean.html.
Courtesy of SOHO consortium. SOHO is a project of
international cooperation between ESA and NASA.]}
\label{soho1}
\end{figure}}

To get
insights regarding the structure of the coronal magnetic field we
have to compute 3D magnetic field models, which use the
measured photospheric magnetic field as the boundary condition. This
procedure is often called ``extrapolation of the coronal magnetic
field from the photosphere.'' In the solar corona the thermal
conductivity is much higher parallel than perpendicular to the field
so that field lines may become visible by the emission at
appropriate temperatures. This makes in some sense magnetic field
lines visible and allows us to test coronal magnetic field models.
In such tests 2D projection of the computed 3D magnetic field
lines are compared with plasma loops seen in coronal images. This
mainly qualitative comparison cannot guarantee that the computed
coronal magnetic field model and derived quantities, like the
magnetic energy, are accurate. Coronal magnetic field lines which
are in reasonable agreement with coronal images are, however, more
likely to reproduce the true nature of the coronal magnetic field.

\epubtkImage{beta}{%
\begin{figure}[htbp]
\centerline{\includegraphics[scale=1.0]{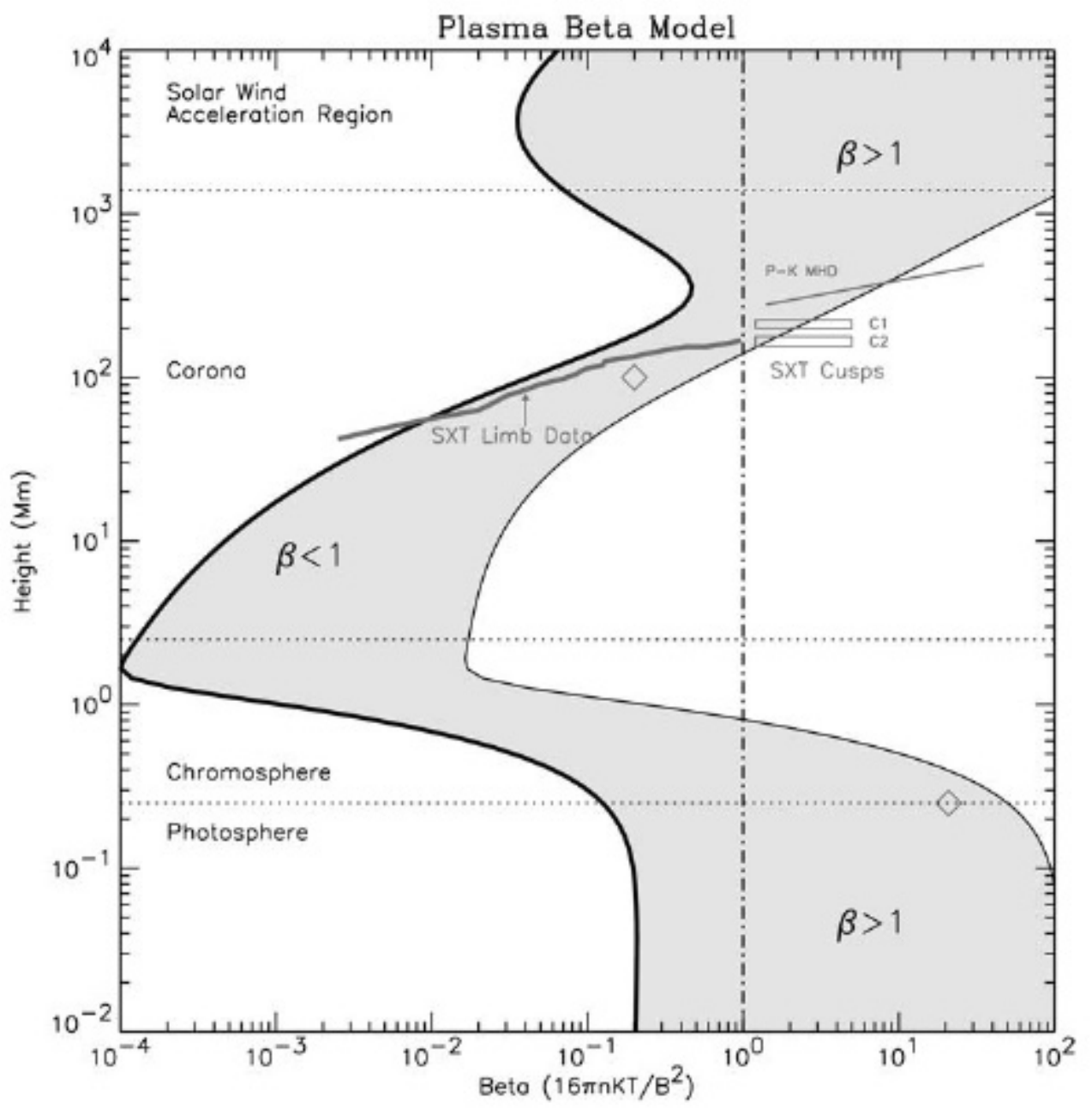}}
\caption{Plasma $\beta$ model over active regions.
The shaded area corresponds to magnetic fields
originating from a sunspot region with 2500~G and a plage region
with 150~G. The left and right boundaries of the shaded area are
related to umbra and plage magnetic field models, respectively.
Atmospheric regions magnetically connected to high magnetic field
strength areas in the photosphere naturally
have a lower plasma $\beta$.
The original figure
was published as Figure~3 in \cite{gary01}.}
\label{figbeta}
\end{figure}}

To model the coronal magnetic field $\mathbf{B}$ we have to introduce
some assumptions. It is therefore necessary to get some a priori
insights regarding the physics of the solar corona. An important
quantity is the plasma $\beta$ value, a dimensionless number
which is defined as the ratio between the plasma pressure $p$
and the magnetic pressure,
\begin{equation}
\beta=2 \mu_0 \frac{p}{B^2}.
\end{equation}
Figure~\ref{figbeta} from \cite{gary01} shows how the plasma
$\beta$ value changes with height in the solar atmosphere.
As one can see a region with $\beta \ll 1$ is sandwiched
between the photosphere and the upper corona, where $\beta$ is
about unity or larger. In regions with $\beta \ll 1$ the
magnetic pressure dominates over the plasma pressure
(and as well over other non-magnetic forces like gravity
and the kinematic plasma flow pressure). Here we can neglect
in the lowest order all non-magnetic forces and assume that
the Lorentz force vanishes. This approach is called the
{\textit{force-free field approximation} and for
static configurations it is defined as:
\begin{eqnarray}
\mathbf{j}\times\mathbf{B} & = & \mathbf{0}, \label{forcebal}\\
\mathbf{j} & = & \frac{1}{\mu_0}\nabla \times \mathbf{B}
\;\; { \rm is \, the \, electric \, current \, density} \label{ampere}, \\
\nabla\cdot\mathbf{B} & = & 0 \label{solenoidal},
\end{eqnarray}
or by inserting Equation~(\ref{ampere}) into (\ref{forcebal}):
\begin{eqnarray}
(\nabla \times \mathbf{B}) \times\mathbf{B} & = & \mathbf{0}, \label{eq:force-free} \\
\nabla\cdot\mathbf{B} & = & 0 .\label{eq:solenoidal}
%
\end{eqnarray}
Equation~(\ref{eq:force-free}) can be fulfilled either by:
%
\begin{equation}
\nabla \times \mathbf{B}=0 \; \quad \mbox{current-free or potential magnetic fields}
\end{equation}
or  by
\begin{equation}
\mathbf{B} \parallel \nabla \times \mathbf{B} \; \quad \mbox{force-free fields}.
\end{equation}

Current free (potential) fields are the simplest assumption for the coronal
magnetic field. The line-of-sight (LOS) photospheric magnetic field which
is routinely measured with magnetographs are used as boundary conditions
to solve the Laplace equation for the scalar potential $\phi$,
%
\begin{equation}
\Delta \phi =0
\end{equation}
where the Laplacian operator $\Delta $ is the divergence of the
gradient of the scalar field and
\begin{equation}
\mathbf{B} = -\nabla \phi .
\end{equation}
%
When one deals with magnetic fields of a global scale, one usually
assumes the so-called ``source surface'' (at about $2.5$ solar
radii where all field lines become radial): See, e.g.,
\citet{schatten69} for details on the potential-field
source-surface (PFSS) model. Figure~\ref{potentialglobal}
shows such a potential-field source-surface model for May
2001 from \cite{wiegelmann:etal04:conf}.

Potential fields are popular due to their mathematical simplicity and
provide a first coarse view of the magnetic structure in the solar corona.
They cannot, however, be used to model the magnetic field in active regions
precisely, because they do not contain free magnetic energy to drive
eruptions. Further, the transverse photospheric magnetic field computed
from the potential-field assumption usually does not agree with measurements
and the resulting potential field lines do deviate from coronal loop
observations. For example, a comparison of global potential fields
with TRACE images by \cite{schrijver:etal05a} and with
stereoscopically-reconstructed loops by \cite{sandman:etal09}
showed large deviations between potential magnetic field lines
and coronal loops.

\epubtkImage{potentialglobal}{%
\begin{figure}[htbp]
\centerline{\includegraphics[scale=0.4]{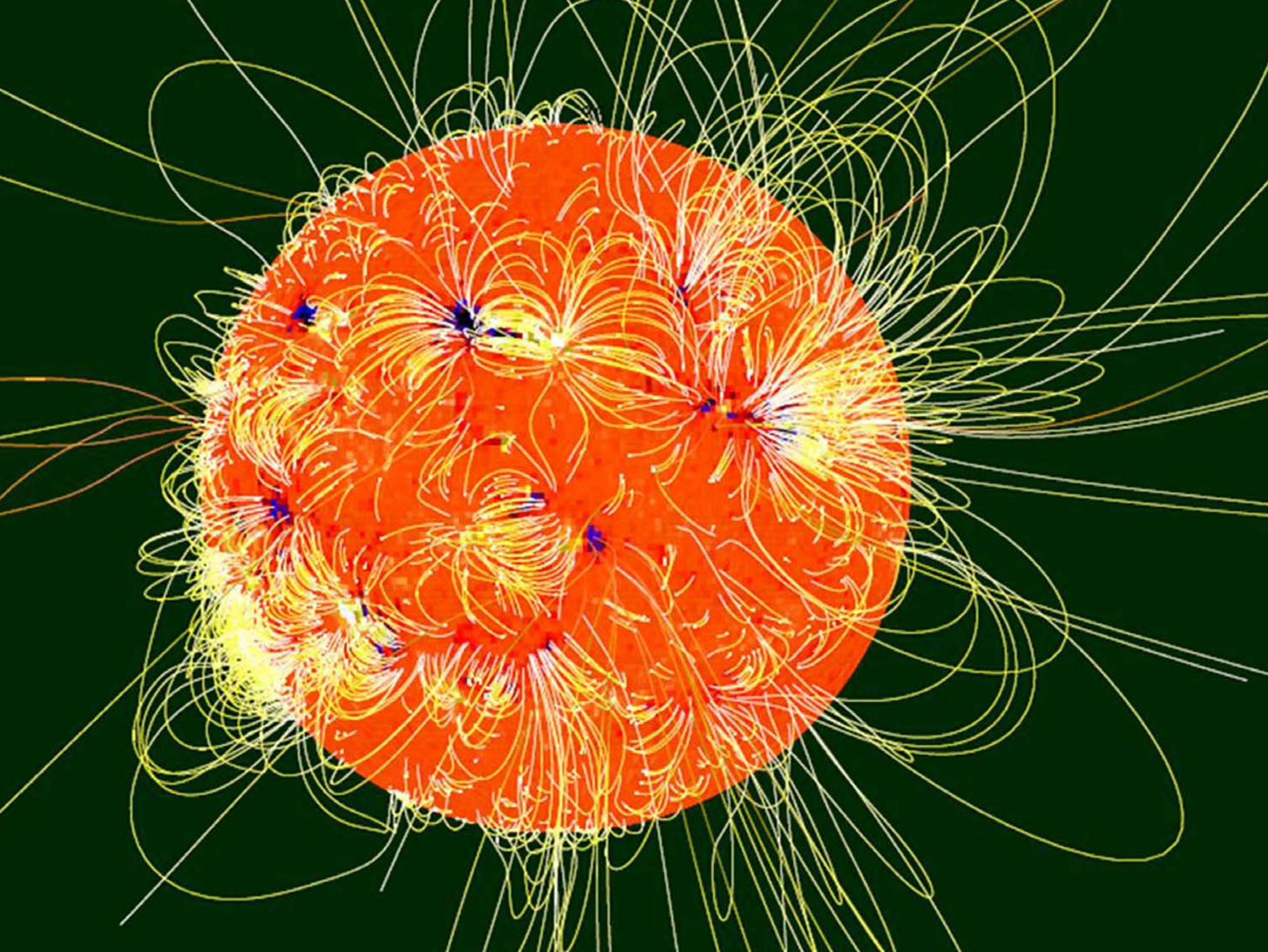}}
\caption{Global potential field reconstruction. The original figure
  was published in \cite{wiegelmann:etal04:conf}.}
\label{potentialglobal}
\end{figure}}

The $\mathbf{B} \parallel \nabla \times \mathbf{B}$ condition can be rewritten as
\begin{eqnarray}
\nabla \times \mathbf{B} & = & \alpha \mathbf{B}
\label{amperealpha}, \\
\mathbf{B} \cdot \nabla \alpha &=& 0
\label{Bgradalpha},
\end{eqnarray}
where $\alpha$ is called the force-free parameter or force-free function.
From the horizontal photospheric magnetic field components $(B_{x0}, \,
B_{y0})$ we can compute the vertical electric current density
\begin{equation}
\mu_0 j_{z0}=\frac{\partial B_{y0}}{\partial x}-\frac{\partial B_{x0}}{\partial y}
\label{j_photo}
\end{equation}
and the corresponding distribution of the force-free function
$\alpha(x,y)$ in the photosphere
\begin{equation}
\alpha(x,y)=\mu_0 \frac{j_{z0}}{B_{z0}} .
\label{alpha0direct}
\end{equation}
Condition~(\ref{Bgradalpha}) has been derived by taking the
divergence of Equation~(\ref{amperealpha}) and using the solenoidal
condition~(\ref{solenoidal}). Mathematically
Equations~(\ref{amperealpha}) and (\ref{Bgradalpha}) are equivalent
to Equations~(\ref{forcebal})\,--\,(\ref{solenoidal}). Parameter
$\alpha$ can be a function of position, but
Equation~(\ref{Bgradalpha}) requires that $\alpha$ be constant along
a field line. If $\alpha$ is constant everywhere in the volume under
consideration, the field is called linear force-free field (LFFF),
otherwise it is nonlinear force-free field (NLFFF).
Equations~(\ref{amperealpha}) and (\ref{Bgradalpha}) constitute
partial differential equations of mixed elliptic and hyperbolic
type. They can be solved  as a well-posed boundary value problem by
prescribing the vertical magnetic field and for one polarity the
distribution of $\alpha$ at the boundaries. As shown by
\cite{bineau72} these boundary conditions ensure the existence and
unique NLFFF solutions at least for small values of $\alpha$ and
weak nonlinearities. \cite{boulmezaoud:etal00} proved the existence
of solutions for a simply and multiply connected domain. As pointed
out by \cite{aly07} these boundary conditions disregard part of the
observed photospheric vector field: In one polarity only the curl of
the horizontal field (Equation (\ref{j_photo})) is used as the
boundary condition, and the horizontal field of the other polarity
is not used at all. For a general introduction to complex boundary
value problems with elliptic and hyperbolic equations we refer to
\cite{kaiser00}.

Please note that high plasma $\beta$ configurations are not necessarily
a contradiction to the force-free condition \citep[see][for
details]{neukirch05}. If the plasma pressure is constant or the
pressure gradient is compensated by the gravity force
$(\nabla p = -\rho \nabla \Psi$,  where $\rho$ is the mass density and
$\Psi$ the gravity potential of the Sun.)
a high-$\beta$ configuration can
still be consistent with a vanishing Lorentz force of the magnetic
field. In this sense a low plasma $\beta$ value is a sufficient,
but not a necessary, criterion for the force-free assumption. In
the generic case, however, high plasma $\beta$ configurations will
not be force-free and the approach of the force-free field is limited
to the upper chromosphere and the corona (up to about $2.5\,R_{\rm s}$).

%
%
\newpage
\section{Linear Force-Free Fields}

Linear force-free fields are characterized by
\begin{eqnarray}
\nabla \times \mathbf{B} & = & \alpha \mathbf{B}, \label{eq11} \\
\nabla \cdot \mathbf{B} & = & 0,
\label{eq12}
\end{eqnarray}
where the force-free parameter $\alpha$ is constant. Taking the
curl of Equation~(\ref{eq11}) and using the solenoidal
condition~(\ref{eq12}) we derive a vector Helmholtz equation:
\begin{equation}
\Delta \mathbf{B} + \alpha^2 \mathbf{B} =0
\label{eq:Helmholtz}
\end{equation}
which can be solved by a separation ansatz, a Green's function method
\citep{chiu:etal77} or a Fourier method \citep{alissandrakis81}.
These methods can also be used to compute a potential field by
choosing $\alpha=0$.

For computing the solar magnetic field in the corona with the
linear force-free model one needs only measurements of the LOS
photospheric magnetic field. The force-free parameter $\alpha$
is a priori unknown and we will discuss later how $\alpha$ can
be approximated from observations. \cite{seehafer78} derived
solutions of the linear force-free equations (local Cartesian
geometry with $(x, y)$ in the photosphere and $z$ is the
height from the Sun's surface) in the form:
\begin{eqnarray}
B_x & = & \sum_{m,n=1}^{\infty} \frac{C_{mn}}{\lambda_{mn}} \exp \left(-r_{mn} z \right) \cdot
\left[ \alpha \frac{\pi n}{L_y} \sin \left(\frac{\pi m x}{L_x}\right)
\cos \left(\frac{\pi n y}{L_y}\right) - \right.
\nonumber \\
&&\left. -r_{mn} \frac{\pi m}{L_x} \cos \left(\frac{\pi m x}{L_x}\right)
\sin \left(\frac{\pi n y}{L_y}\right)
\right] ,
\label{B_x}
\\
B_y & = & -\sum_{m,n=1}^{\infty} \frac{C_{mn}}{\lambda_{mn}} \exp \left(-r_{mn} z \right) \cdot
\left[ \alpha \frac{\pi m}{L_x} \cos \left(\frac{\pi m x}{L_x}\right)
\sin \left(\frac{\pi n y}{L_y}\right)
+ \right. \nonumber \\
&& \left.+r_{mn} \frac{\pi n}{L_y} \sin \left(\frac{\pi m x}{L_x}\right)
\cos \left(\frac{\pi n y}{L_y}\right)
\right] , \\
B_z & = & \sum_{m,n=1}^{\infty} C_{mn} \exp \left(-r_{mn} z \right) \cdot
\sin \left(\frac{\pi m x}{L_x}\right)
\sin \left(\frac{\pi n y}{L_y}\right) ,
\label{B_z}
\end{eqnarray}
with $\lambda_{mn}= \pi^2 (m^2/L_x^2+ n^2/L_y^2 )$ and $r_{mn}=\sqrt{\lambda_{mn}-\alpha^2}$.

As the boundary condition, the method uses the distribution of
$B_z(x,y)$ on the photosphere $z=0$. The coefficients $C_{mn}$
can be obtained by comparing Equation~(\ref{B_z}) for $z=0$
with the magnetogram data. In practice Seehafer's
(\citeyear{seehafer78}) method is used for calculating
the linear force-free field (or potential field for $\alpha=0$)
for a given magnetogram (e.g., MDI on SOHO) and a given value of
$\alpha$ as follows. The observed magnetogram which covers a
rectangular region extending from $0$ to $L_x$ in $x$ and
$0$ to $L_y$ in $y$ is artificially extended onto a rectangular
region covering $-L_x$ to $L_x$ and $-L_y$ to $L_y$ by taking
an antisymmetric mirror image of the original magnetogram in
the extended region, i.e.,
\begin{eqnarray*}
B_z(-x,y) &= &-B_z(x,y), \\
B_z(x,-y) &= &-B_z(x,y), \\
B_z(-x,-y) &=& B_z(x,y) \qquad (0<x<L_x, 0<y<L_y).
\end{eqnarray*}
This makes the total magnetic flux in the whole extended region to be zero.
(Alternatively one may pad the extended region with zeros, although
in this case the total magnetic flux may be non-zero.) The coefficients
$C_{mn}$ are derived from this enlarged magnetogram with the help of a
Fast Fourier Transform. In order for  $r_{mn}$ to be real and
positive so that solutions (\ref{B_x})--(\ref{B_z}) do not diverge
at infinity, $\alpha^2$ should not exceed the maximum value for
given $L_x$ and $L_y$,
\[
\alpha^2_{\mbox{max}}= \pi^2\left(\frac{1}{L_x^2} +\frac{1}{L_y^2}\right).
\]
Usually $\alpha$ is normalized by the harmonic mean $L$ of $L_x$ and $L_y$ defined by
\[
\frac{1}{L^2}=\frac{1}{2}\left( \frac{1}{L_x^2}+\frac{1}{L_y^2}\right) .
\]
For $L_x=L_y$ we have $L=L_x=L_y$. With this normalization
the values of $\alpha$ fall into the range $-\sqrt{2} \pi < \alpha < \sqrt{2} \pi$.

\epubtkImage{lauraall}{%
\begin{figure}[htbp]
\centerline{\includegraphics[scale=0.7]{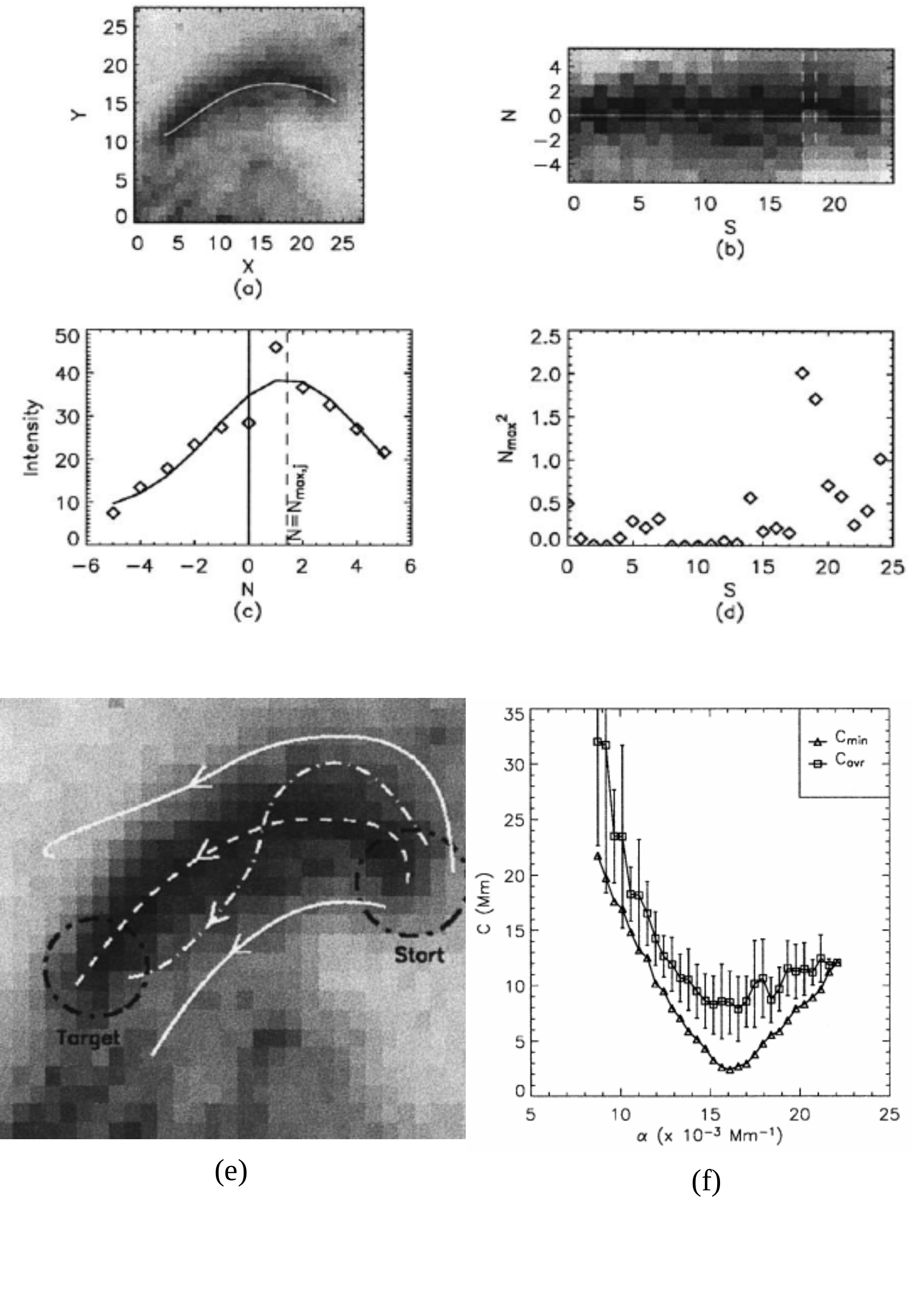}}
\caption{How to obtain the optimal linear force-free parameter $\alpha$
from coronal observations. [The original figures were published
as Figures~3, 4, and 5 in \cite{carcedo:etal03}].}
\label{laura1}
\end{figure}}

\subsection{How to obtain the force-free parameter $\alpha$}

Linear force-free fields require the LOS magnetic field in the photosphere as
input and contain a free parameter $\alpha$. One possibility to approximate
$\alpha$ is to compute an averaged value of $\alpha$ from the measured
horizontal photospheric magnetic fields as done, e.g., in
\cite{pevtsov:etal94}, \cite{wheatland99}, \cite{leka:etal99}, and
\cite{hagino:etal04}, where \cite{hagino:etal04} calculated an averaged value
$\alpha=\sum {\mu_0 J_z {\rm sign}(B_z)}/ \sum{|B_z|}$. The vertical electric
current in the photosphere is computed from the horizontal photospheric field
as $J_z=\frac{1}{\mu_0}\left(\frac{\partial B_y}{\partial x}-\frac{\partial
B_x}{\partial y}\right)$. Such approaches derive best fits of a linear
force-free parameter $\alpha$ with the measured horizontal photospheric
magnetic field.

Alternative methods use coronal observations to find the optimal
value of $\alpha$. This approach usually means that one computes
several magnetic field configurations with varying values of
$\alpha$ in the allowed range and to compute the corresponding
magnetic field lines. The field lines are then projected onto
coronal plasma images. A method developed by
\cite{carcedo:etal03} is shown in Figure~\ref{laura1}. In this
approach the shape of a number of field lines with different values
of $\alpha$, which connect the foot point areas (marked as start and
target in Figure~\ref{laura1}(e)) are compared with a coronal image.
For a convenient quantitative comparison the original image shown in
Figure~\ref{laura1}(a) is converted to a coordinate system using the
distances along and perpendicular to the field line, as shown in
Figure~\ref{laura1}(b). For a certain number of $N$ points along
this uncurled loop the perpendicular intensity profile of the
emitting plasma is fitted by a Gaussian profile in
Figure~\ref{laura1}(c) and the deviation between field line and
loops are measured in Figure~\ref{laura1}(d). Finally, the optimal
linear force-free value of $\alpha$ is obtained by minimizing this
deviation with respect to $\alpha$, as seen in
Figure~\ref{laura1}(f).

The method of \cite{carcedo:etal03} has been developed mainly with
the aim of computing the optimal $\alpha$ for an individual coronal
loop and involves several human steps, e.g., identifying an
individual loop and its footpoint areas and it is required that the
full loop, including both footpoints, is visible. This makes it
somewhat difficult to apply the method to images with a large number
of loops and when only parts of the loops are visible. For EUV loops
it is also often not possible to identify both footpoints. These
shortcomings can be overcome by using feature recognition
techniques, e.g., as developed in \cite{aschwanden:etal08} and
\cite{inhester:etal08} to extract one-dimensional curve-like
structures (loops) automatically out of coronal plasma images. These
identified loops can then be directly compared with the projections
of the magnetic field lines, e.g., by computing the area spanned
between the loop and the field line as defined in
\cite{wiegelmann:etal06}. This method has become popular in
particular after the launch of the two STEREO spacecraft in October
2006 \citep{kaiser:etal08}. The projections of the 3D linear
force-free magnetic field lines can be compared with images from two
vantage viewpoints as done for example in
\cite{feng:etal07,feng:etal07a}. This automatic method applied to a
number of loops in one active region revealed, however, a severe
shortcoming of linear force-free field models. The optimal linear
force-free parameter $\alpha$ varied for different field lines,
which is a contradiction to the assumption of a linear model. A
similar result was obtained by \cite{wiegelmann:etal02} who tried to
fit the loops stereoscopically reconstructed by
\cite{aschwanden:etal99}. On the other hand, \cite{marsch:etal04}
found in their example that one value of $\alpha$ was sufficient to
fit several coronal loops. Therefore, the fitting procedure tells us
also whether an active region can be described consistently by a
linear force-free field model: Only if the scatter in the optimal
$\alpha$ values among field lines is small, one has a consistent
linear force-free field model which fits coronal structures.  In
the generic case that $\alpha$ changes significantly between field
lines, one cannot obtain a self-consistent force-free field by a
superposition of linear force-free fields, because the resulting
configurations are not force-free. As pointed out by
\cite{malanushenko:etal09} it is possible, however, to estimate
quantities like twist and loop heights with an error about of 15\%
and 5\%, respectively. The price one has to pay is using a model
that is not self-consistent.

\epubtkImage{beta}{%
\begin{figure}[htbp]
\mbox{\includegraphics[width=7cm,height=6cm]{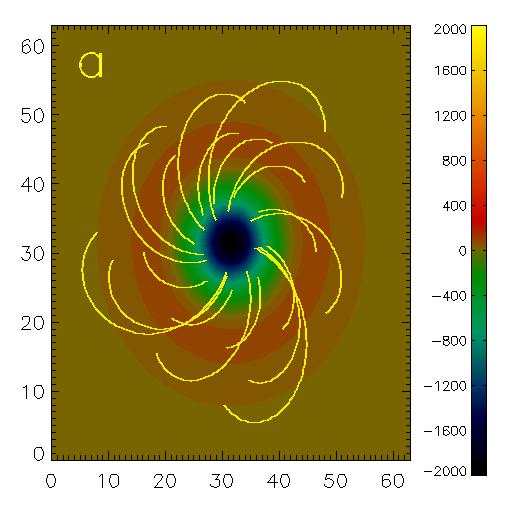}
\includegraphics[width=7cm,height=6cm]{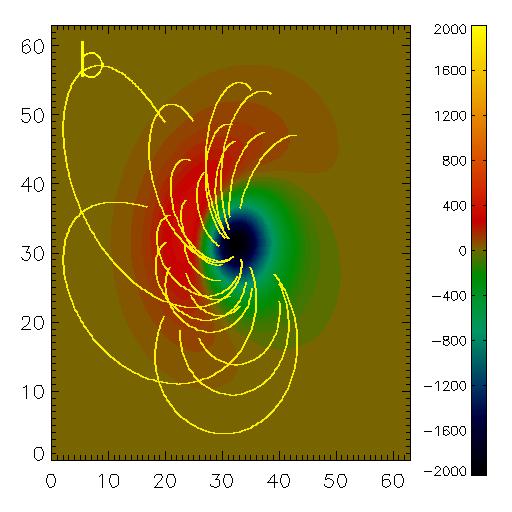}}
\mbox{\includegraphics[width=7cm,height=6cm]{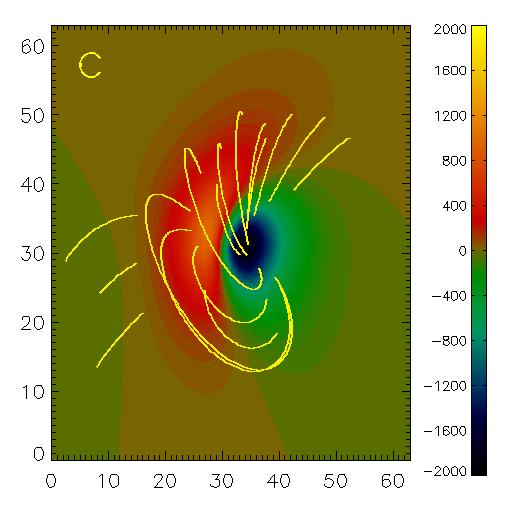}
\includegraphics[width=7cm,height=6cm]{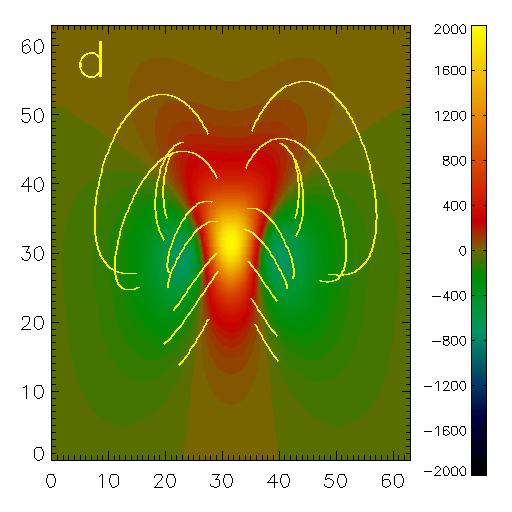}}
\caption{Low and Lou's (\citeyear{low:etal90}) analytic nonlinear
  force-free equilibrium. The original 2D equilibrium is invariant in
  $\varphi$, as shown in panel a. Rotating the 2D-equilibrium and a
  transformation to Cartesian coordinates make this symmetry less
  obvious (panels b--d), where the equilibrium has been rotated by an
  angle of $\varphi=\frac{\pi}{8}, \, \frac{\pi}{4}$, and
  $\frac{\pi}{2}$, respectively.
  The colour-coding corresponds to the vertical magnetic field
  strength in G (gauss) in the photosphere ($z=0$ in the model) and a number
  of arbitrary selected magnetic field lines are shown in yellow. }
\label{figlowlou}
\end{figure}}

%
%
\newpage
\section{Analytic or Semi-Analytic Approaches to Nonlinear Force-Free Fields}
\label{NLFFF-2D}
Solving the nonlinear force-free equations in full 3-D
is extremely difficult.
Configurations with one or two invariant coordinate(s) are more suitable
for an analytic or
semi-analytic treatment. Solutions in the form of an infinitely long cylinder
with axial symmetry are the simplest cases, and two best known examples are
Lundquist's (\citeyear{lundquist50}) solution in terms of Bessel functions
($\alpha$=constant), and a solution used by \citet{gold-hoyle60} in their
flare model ($\alpha \neq$ constant, all field lines have the same pitch in
the direction of the axis). \citet{low73} considered a 1D Cartesian (slab)
geometry and analyzed slow time evolution of the force-free field with
resistive diffusion.

In Cartesian 2D geometry with one ignorable coordinate in the
horizontal (depth) direction, one ends up with a second-order
partial differential equation, called the Grad--Shafranov equation
in plasma physics. The force-free Grad--Shafranov equation is a
special case of the Grad--Shafranov equation for magneto-static
equilibria \citep[see][]{grad:etal58}, which allow to compute
plasma equilibria with one
ignorable coordinate, e.g. a translational, rotational or helical
symmetry. For an overview on how the Grad--Shafranov equation can be
derived for arbitrary curvilinear coordinates with axisymmetry we
refer to \citep[][section 3.2.]{marsh:96}. In the cartesian case one
finds \cite[see, e.g.,][section 13.4]{sturrock94}
\begin{equation}
\Delta A = -\lambda^2 \, f(A)
\end{equation}
where the magnetic flux function $A$ depends only on two spatial
coordinates and any choice of $f(A)$ generates a solution of a
magneto-static equilibrium with symmetry. For static equilibria with
a vanishing plasma pressure gradient the method naturally provides
us force-free configurations. A popular choice for the generating
function is an exponential ansatz, see, e.g.
\citeauthor{low77}(\citeyear{low77}),
\citeauthor{birn:etal78}(\citeyear{birn:etal78}),
\citeauthor{priest:etal80}(\citeyear{priest:etal80}). The existence
of solutions (sometimes multiple, sometimes none) and bifurcation of
a solution sequence have been extensively investigated
\citep[e.g.,][]{birn-schindler81}. We will consider the
Grad--Shafranov equation in spherical polar coordinates in the
following.

\subsection{Low and Lou's (1990) equilibrium}
As an example we refer to \cite{low:etal90}, who solved the Grad--Shafranov
equation in spherical coordinates $(r, \theta, \varphi)$ for axisymmetric
(invariant in $\varphi$) nonlinear force-free fields. In this case the
magnetic field is assumed to be written in the form
\begin{equation}
\mathbf{B} = \frac{1}{r \sin\theta} \; \left(
\frac{1}{r} \, \frac{\partial A}{\partial \theta} \mathbf{e}_r-
\frac{\partial A}{\partial r} \mathbf{e}_{\theta} +
Q \, \mathbf{e}_{\varphi} \right) ,
\label{lowloub}
\end{equation}
where $A$ is the flux function, and $Q$ represents the $\varphi$-component of
the magnetic field $\mathbf{B}$, which depends only on $A$. This ansatz
automatically satisfies the solenoidal condition~(\ref{eq:solenoidal}), and
the force-free equation~(\ref{eq:force-free}) reduces to a Grad--Shafranov
equation for the flux function $A$
\begin{equation}
\frac{\partial^2 A}{\partial r^2}+\frac{1-\mu^2}{r^2} \,
\frac{\partial^2 A}{\partial \mu^2}+Q \; \frac{d \, Q}{d \, A} =0 ,
\label{GS}
\end{equation}
where $\mu=\cos\theta$. \citet{low:etal90} looked for solutions in the form
\begin{equation}
Q(A) = \lambda A^{1+1/n}  \qquad (\alpha = \frac{d Q}{d A}  \sim A^{1/n})
\end{equation}
with a separation ansatz
\begin{equation}
A(r,\theta)=\frac{P(\mu)}{r^n}.
\end{equation}
Here $n$ and $\lambda$ are constants and $n$ is not necessarily an integer;
$n=1$ and $\lambda=0$ corresponds to a dipole field. Then Equation~(\ref{GS})
reduces to an ordinary differential equation for $P(\mu)$, which can be
solved numerically. Either by specifying $n$ or $\lambda$, the other is
determined as an eigenvalue problem \citep{wolfson95}.The solution in 3D
space is axisymmetric and has a point source at the origin. This symmetry is
also visible after a transformation to Cartesian geometry as shown in
Figure~\ref{figlowlou}(a). The symmetry becomes less obvious, however,
when the symmetry axis is rotated with respect to the Cartesian
coordinate axis; see Figures~\ref{figlowlou}(b)--(d). The resulting
configurations are very popular for testing numerical algorithms for a
3D NLFFF modeling. For such tests the magnetic field vector on the
bottom boundary of a computational box is extracted from the
semi-analytic Low-Lou solution and used as the boundary condition for
numerical force-free extrapolations. The quality of the reconstructed
field is evaluated by quantitative comparison with the exact solution;
see, e.g., \cite{schrijver:etal06}. Similarly one can shift the origin
of the point source with respect to the Sun center and the solution is
not symmetric to the Sun's surface and can be used to test spherical codes.

\subsection{Titov--D\'emoulin equilibrium}

Another approach for computing axisymmetric NLFFF solutions has been
developed in \cite{titov:etal99}. This model active region contains
a current-carrying flux-tube, which is imbedded into a
potential field. A motivation for such an approach is that solar
active regions may be thought of as composed of such flux tubes. The
method allows to study a sequence of force-free configurations
through which the flux tube emerges. Figure~\ref{td1} shows how the
equilibrium is built up. The model contains a symmetry axis, which
is located at a distance $d$ below the photosphere. A line current
$I_0$ runs along this symmetry axis and creates a circular potential
magnetic field. This potential field becomes disturbed by a toroidal
ring current $I$ with the minor radius $a$ and the major radius $R$,
where $a \ll R$ is assumed. Two opposite magnetic monopoles of
strength $q$ are placed on the axis separated by distance $L$. These
monopoles are responsible for the poloidal potential field. This
field has its field lines overlying the force-free current and
stabilizes the otherwise unstable configuration. Depending on
the choice of parameters one can contain stable or unstable
nonlinear force-free configurations. The unstable branch of this
equilibrium has been used to study the onset of coronal mass
ejections; see Section~\ref{numerical_stability}. Stable branches of
the Titov--D\'emoulin equilibrium are used as a challenging test for
numerical NLFFF extrapolation codes \citep[see,
e.g.,][]{wiegelmann:etal06a,valori:etal10}.

\epubtkImage{td1}{%
\begin{figure}[htbp]
\centerline{\includegraphics[width=0.5\textwidth]{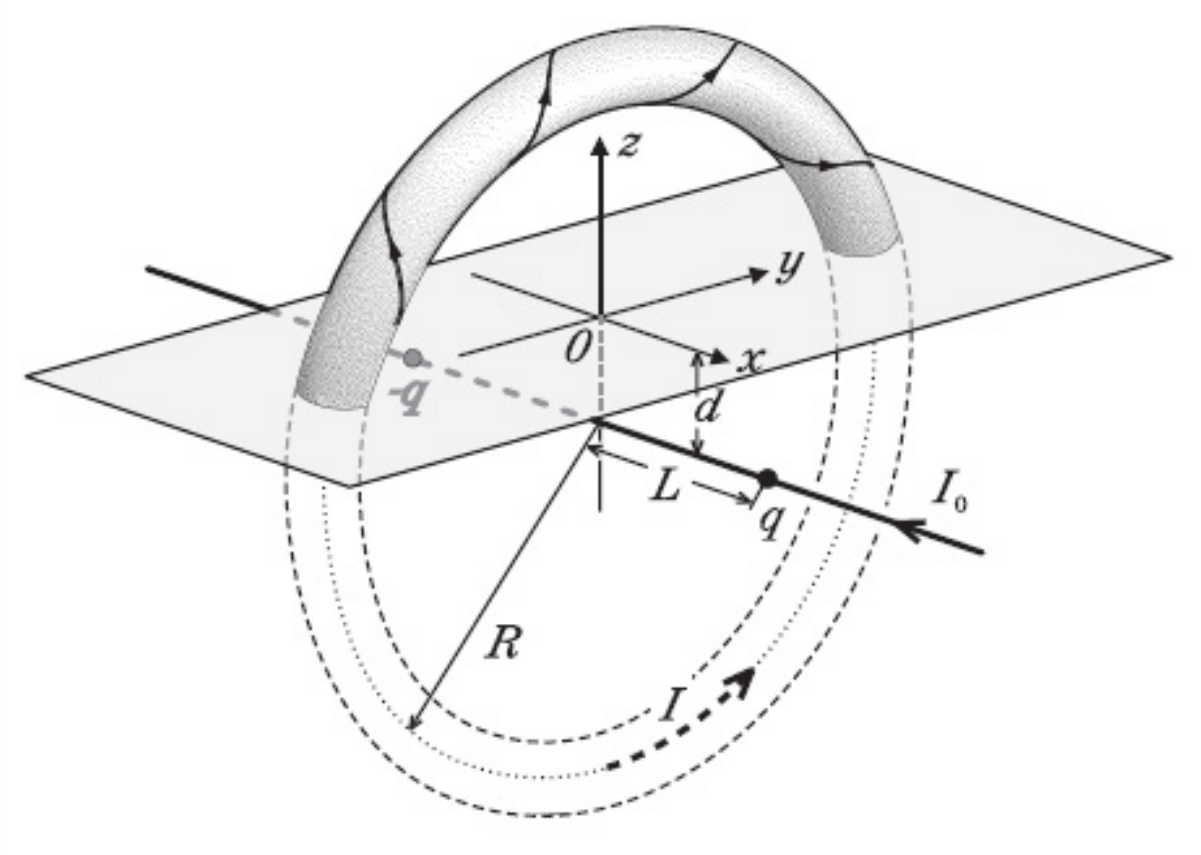}}
\caption{Construction of the Titov--D\'emoulin equilibrium. The original figure
was published as Figure~2 in \cite{titov:etal99}.}
\label{td1}
\end{figure}}

\newpage
\section{Azimuth Ambiguity Removal and Consistency of Field Measurements}
\label{ambiguity}
\subsection{How to derive vector magnetograms?}
\label{inversion_code}
NLFFF extrapolations require the photospheric magnetic field vector as input.
Before discussing how this vector can be extrapolated into the solar
atmosphere, we will address known problems regarding the photospheric field
measurements. Vector magnetographs are being operated daily at NAOJ/Mitaka
\citep{sakurai:etal95}, NAOC/Huairou \citep{ai:etal86}, NASA/MSFC
\citep{hagyard:etal82}, NSO/Kitt Peak \citep{henney06}, and U.~Hawaii/Mees
Observatory \citep{mickey:etal96}, among others. The Solar Optical Telescope
\citep[SOT;][]{tsuneta:etal08} on the Hinode mission has been taking vector
magnetograms since 2006. Full-disk vector magnetograms are observed routinely
since 2010 by the Helioseismic and Magnetic Imager (HMI;
\citet{scherrer:etal12}) onboard the Solar Dynamics Observatory (SDO).
Measurements with these vector magnetographs provide us eventually with the
magnetic field vector on the photosphere, say $B_{z0}$ for the vertical and
$B_{x0}$ and $B_{y0}$ for the horizontal fields.
Deriving these quantities from measurements is an involved
physical process} based on the Zeeman and Hanle effects and the
related inversion of Stokes profiles
\citep[e.g.,][]{labonte:etal99}. Within this work we only
outline the main steps and refer to
 \citeauthor{iniesta:etal96}(\citeyear{iniesta:etal96}),
 \citeauthor{iniesta:etal03}(\citeyear{iniesta:etal03}), and
 \citeauthor{landi94}(\citeyear{landi94})
for details.
Actually measured are polarization degrees across magnetically
sensitive spectral lines, e.g. the line pair Fe {\sc i} $6302.5$ and
$6301.5$ {\AA} as used on Hinode/SOT \citep[see][]{lites:etal07} or
Fe {\sc i} $6173.3$ {\AA} as used on SDO/HMI
\citep[see][]{schou:etal12}. The accuracy of these measurements
depends on the spectral resolution, for example the HMI instruments
measures at six points in the Fe {\sc i} $6173.3$ {\AA} absorption
line. In a subsequent step the Stokes profiles are inverted to
derive the magnetic field strength, its inclination and azimuth. One
possibility to carry out the inversion \citep[see][]{lagg:etal04} is
to fit the measured Stokes profiles with synthetic ones derived from
the Unno-Rachkovsky solutions \citep{unno:56,rachkowsky:67}. Usually
one assumes a simple radiative transfer model like the
Milne-Eddington atmosphere \citep[see e.g.][]{landi92} in order to
derive the analytic Unno-Rachkovsky solutions. The line-of-sight
component of the field is approximately derived by $B_\ell \propto
V/I$, where $V$ is the circular polarization and $I$ the intensity
 (the so-called weak-field approximation).
The error from photon noise is approximately $\delta B_\ell \propto
\frac{\delta V}{I}$, where $\delta$ corresponds to noise in the
measured and derived quantities. As a rule of thumb, $\delta V/I
\sim 10^{-3}$ and $\delta B_\ell \sim$ a few gauss (G) in currently
operating magnetographs. The horizontal field components can be
approximately derived from the linear polarization $Q$ and $U$ as
$B_t ^2 \propto \sqrt{Q^2 +U^2}/I$. The error in $\delta B_t$ is
estimated as $ B_t \, \delta B_t \propto \frac{Q \delta Q+U \delta
U}{\sqrt{Q^2+U^2} I}$ from which the minimum detectable $B_t \,
(\delta B_t \sim B_t$) is proportional to the square root of the
photon noise $\approx \sqrt{\delta Q^2 +\delta U^2}/I \approx
\sqrt{\delta V/I}$, namely around a few tens of G, one order of
magnitude higher than $\delta B_\ell$. (Although $\delta B_t$ scales
as $1/B_t$ and gives much smaller $\delta B_t$ for stronger $B_t$,
one usually assumes a conservative error estimate that $\delta B_t
\sim$ a few tens of G regardless of the magnitude of $B_t$.)

Additional complications occur when the observed region is far away
from the disk center and consequently the line-of-sight and vertical
magnetic field components are far apart \citep[see][for
  details]{gary:etal90}. The inverted horizontal magnetic field
components $B_{x0}$ and $B_{y0}$ cannot be uniquely derived, but
contain a $180^\circ$ ambiguity in azimuth, which has to be removed
before the fields can be extrapolated into the corona. In the
following we will discuss this problem briefly. For a more detailed
review and a comparison and performance check of currently available
ambiguity-removal routines with synthetic data, see \cite{metcalf:etal06}.

To remove the ambiguity from this kind of data, some a priori assumptions
regarding the structure of the magnetic field vector are necessary, e.g.,
regarding smoothness. Some methods require also an approximation regarding
the 3D magnetic field structure (usually from a potential field
extrapolation); for example to minimize the divergence of magnetic field
vector or the angle with respect to the potential field. We are mainly
interested here in automatic methods, although manual methods are also
popular, e.g., the AZAM code. If we have in mind, however, the huge data
stream from SDO/HMI, fully automatic methods are desirable. In the following
we will give a brief overview on the ambiguity removal techniques and tests
with synthetic data.

\subsection{Quantitative comparison of ambiguity removal algorithms}

\cite{metcalf:etal06} compared several algorithms and implementations
quantitatively with the help of two synthetic data sets, a flux-rope
simulation by \cite{fan:etal04} and a multipolar constant-$\alpha$ structure
computed with the \cite{chiu:etal77} linear force-free code. The results of
the different ambiguity removal techniques have been compared with a number
of metrics \citep[see Table~II in][]{metcalf:etal06}. For the discussion here
we concentrate only on the first test case (flux rope) and the area metrics,
which simply tells for what fraction of pixels the ambiguity has been removed
correctly. A value of 1 corresponds to a perfect result and 0.5 to
random. The result is visualized in Figure~\ref{ambi1}, where the ambiguity
has been removed correctly in black areas. Wrong pixels are white. In the
following we briefly describe the basic features of these methods and provide
the performance (fraction of pixels with correctly removed ambiguity).

\epubtkImage{ambi1}{%
\begin{figure}[htbp]
\centerline{\includegraphics[scale=0.7]{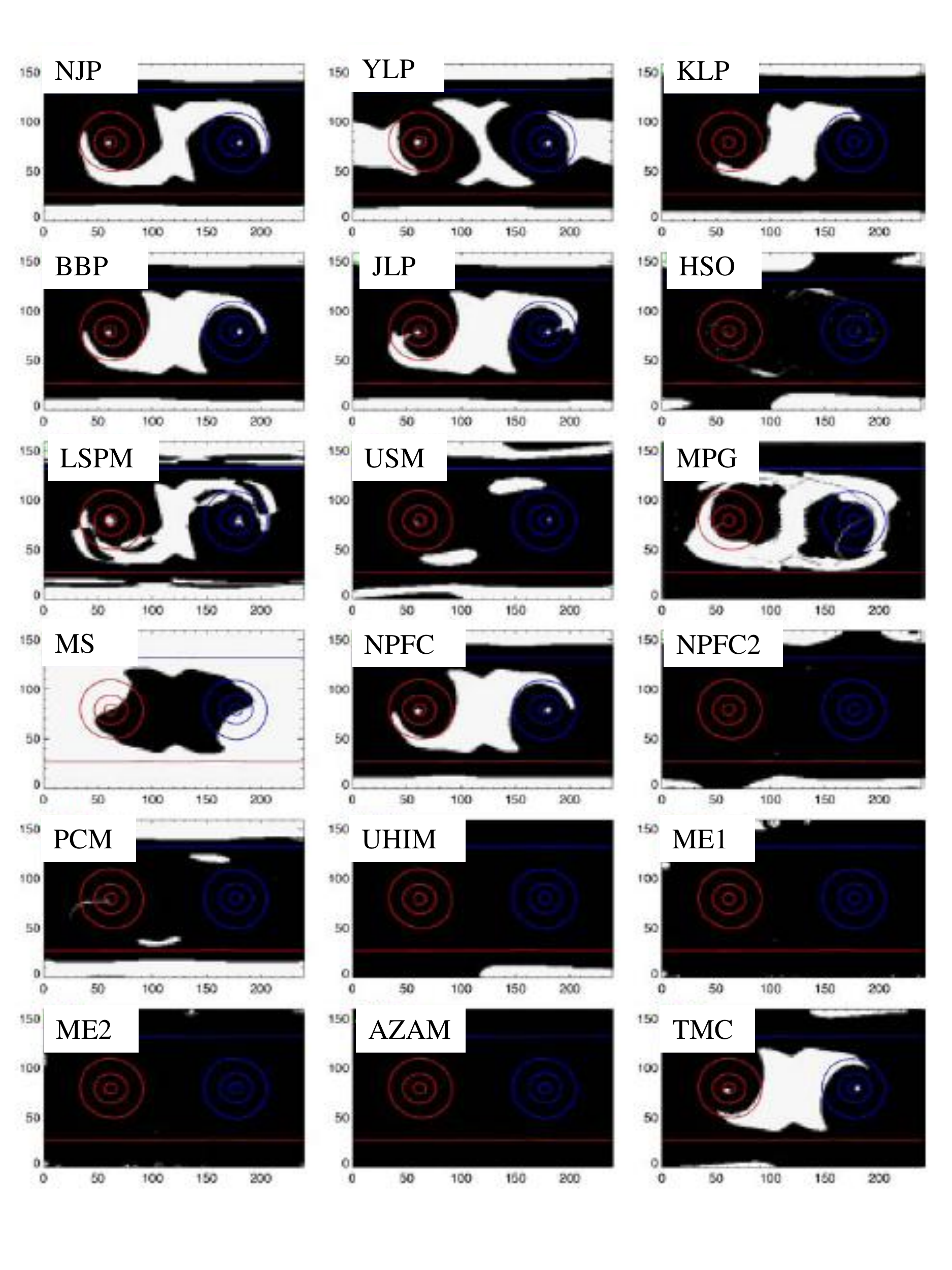}}
\caption{Overview of the performance of different algorithms for
removing the $180^\circ$ azimuth ambiguity. The codes have been
applied to synthetic data (a flux-rope simulation by \cite{fan:etal04}).
In black areas the codes found the correct azimuth and in white areas not.
The original figure was published as Figure~3 in \cite{metcalf:etal06}.}
\label{ambi1}
\end{figure}}

\subsection{Ambiguity removal algorithm}

\subsubsection{Acute angle method}

The magnetic field in the photosphere is usually not force-free and even not
current-free, but an often made assumption is that from two possible
directions ($180^\circ$ apart) of the observed field $\mathbf{B}^{\rm obs}$, the
solution with the smaller angle to the potential field (or another suitable
reference field) $\mathbf{B}^{0}$ is the more likely candidate for the true
field. Consequently we get for the horizontal/transverse\epubtkFootnote{In the
following we assume observations close to the disk center for simplicity,
when the vertical and LOS-component are identical. For observations far away
from the disk center one has to resolve first the ambiguity and apply
coordinate transformations from LOS/transverse to vertical/horizontal fields
afterwards.} field components $\mathbf{B}_t$ the condition
\begin{equation}
\mathbf{B}_t^{\rm obs} \cdot \mathbf{B}_t^{0} >0.
\end{equation}
This condition is easy to implement and fast in application. In
\cite{metcalf:etal06} several different implementations of the acute angle
method are described, which mainly differ by the algorithms used to compute
the reference field. The different implementations of the acute angle methods
got a fraction of 0.64\,--\,0.75 pixels correct (see Figure~\ref{ambi1}, panels
marked with NJP, YLP, KLP, BBP, JLP, and LSPM).

\subsubsection{Improved acute angle methods}
\label{improved_acute_angle}

A sophistication of the acute angle method uses linear force-free fields
\citep{wang97,wang:etal01}, where the optimal force-free parameter $\alpha$
is chosen to maximize the integral
\begin{equation}
S=\int
\frac{|\mathbf{B}^{\rm obs} \cdot \mathbf{B}^{\rm lff}|}{B^{\rm obs} \, B^{\rm lff}}
dx \, dy
\end{equation}
where $\mathbf{B}^{\rm lff}$ is the linear force-free reference field. A
fraction of 0.87 pixels has been identified correctly (see
Figure~\ref{ambi1} second row, right panel marked with HSO).

Another approach, dubbed uniform shear method by \cite{moon:etal03} uses the
acute angle method (with a potential field as reference) only as a first
approximation and subsequently uses this result to estimate a uniform shear
angle between the observed field and the potential field. Then the acute
angle method is applied again to resolve the ambiguity, taking into account
the average shear angle between the observed field and the calculated
potential field. A fraction of 0.83 pixels has been identified correctly.
Consequently both methods significantly improve the potential-field acute
angle method (see Figure~\ref{ambi1} third row, center panel marked with
USM).

\subsubsection{Magnetic pressure gradient}

The magnetic pressure gradient method \citep{cuperman:etal93} assumes a
force-free field and that the magnetic pressure $B^2/2$ decreases with
height. Using the solenoidal and force-free conditions, we can compute the
vertical magnetic pressure gradient as:
\begin{equation}
\frac{1}{2} \frac{\partial B^2}{\partial z}=
B_x \frac{\partial B_z}{\partial x}+
B_y \frac{\partial B_z}{\partial y} -
B_z \left(\frac{\partial B_x}{\partial x}+\frac{\partial B_y}{\partial y} \right)
\end{equation}
with any initial choice for the ambiguity of the horizontal magnetic field
components $(B_x, B_y)$. Different solutions of the ambiguity removal method
give the same amplitude, but opposite sign for the vertical pressure
gradient. If the vertical gradient becomes positive, then the transverse
field vector is reversed. For the test this method got a fraction of 0.74
pixels correct, which is comparable with the potential-field acute angle
method (see Figure~\ref{ambi1} forth row, left panel marked with MS).

\subsubsection{Structure minimization method}

The structure minimization method \citep{georgoulis:etal04} is a
semi-analytic method which aims at eliminating dependencies between pixels.
We do not describe the method here, because in the test only for a fraction
of $0.22$ pixels the ambiguity has been removed correctly, which is worse
than a random result (see Figure~\ref{ambi1} third row, right panel marked
with MPG).

\subsubsection{Non-potential magnetic field calculation method}

The non-potential magnetic field method developed by \cite{georgoulis05} is
identical with the acute angle method close to the disk center. Away from the
disk center the method is more sophisticated and uses the fact that the
magnetic field can be represented as a combination of a potential field and a
non-potential part $\mathbf{B}=\mathbf{B}_{\rm p}+\mathbf{B}_{\rm c}$, where the
non-potential part $\mathbf{B}_{\rm c}$ is horizontal on the boundary and only
$\mathbf{B}_{\rm c}$ contains electric currents. The method aims at computing a
fair a priori approximation of the electric current density before the
ambiguity removal. With the help of a Fourier method the component
$\mathbf{B}_{\rm c}$ and the corresponding approximate field
$\mathbf{B}$ are computed. This field is then used as the reference
field for an acute angle method. The quality of the reference field
depends on the accuracy of the a priori assumed electric current
density $j_z$. In the original implementation by \cite{georgoulis05}
$j_z$ was chosen once a priori and not changed afterwards. In an
improved implementation (published as part of the comparison paper by
\cite{metcalf:etal06} and implemented by Georgoulis) $j_z$ becomes
updated in an iterative process. The original implementation got 0.70
pixels correct and the improved version 0.90 (see Figure~\ref{ambi1}
forth row, center and right panels marked with NPFC and NPFC2,
respectively). So the original method is on the same level as the
potential-field acute angle method, but the current iteration
introduced in the updated method gives significantly better
results. This method has been used for example to resolve the
ambiguity of full-disk vector magnetograms from the SOLIS instrument
\citep{henney06} at NSO/Kitt Peak.

\subsubsection{Pseudo-current method}

The pseudo-current method developed by \cite{gary:etal95} uses as the initial
step the potential-field acute angle method and subsequently applies this
result to compute an approximation for the vertical electric current density.
The current density is then approximated by a number of local maxima of $j_z$
with an analytic expression containing free model parameters, which are
computed by minimizing a functional of the square of the vertical current
density. This optimized current density is then used to compute a correction
to the potential field. This new reference field is then used in the acute
angle method to resolve the ambiguity. In the test case this method got a
fraction of 0.78 of pixels correct, which is only slightly better than the
potential-field acute angle method (see Figure~\ref{ambi1} fifth row, left
panel marked with PCM).

\subsubsection{U. Hawai'i iterative method}
\label{UHIM}

This method, originally developed in \cite{canfield:etal93} and subsequently
improved by a group of people at the Institute for Astronomy, U.~Hawai'i. As
the initial step the acute angle method is applied, which is then improved by
a constant-$\alpha$ force-free field, where $\alpha$ has to be specified by
the user (in principle it should also be possible to apply an automatic
$\alpha$-fitting method as discussed in Section~\ref{improved_acute_angle}).
Therefore, the result would be similar to the improved acute angle methods,
but additional two more steps have been introduced for a further improvement.
In a subsequent step the solution is smoothed (minimizing the angle between
neighboring pixels) by starting at a location where the field is radial and
the ambiguity is obvious, e.g., the umbra of a sunspot. Finally also the
magnetic field divergence or vertical electric current density is minimized.
This code includes several parameters, which have to be specified by the
user. In the test case the code recognized a fraction of 0.97~pixels
correctly. So the additional steps beyond the improved acute angle method
provide another significant improvement and almost the entire region has been
correctly identified (see Figure~\ref{ambi1} fifth row, center panel marked
with UHIM).

\subsubsection{Minimum energy methods}

The minimum energy method has been developed by \cite{metcalf94}. As other
sophisticated methods it uses the potential-field acute angle method as the
initial step. Subsequently a pseudo energy, which is defined as a combination
of the magnetic field divergence and electric current density, is minimized.
In the original formulation the energy was defined as
$E=\sum (|\nabla \cdot \mathbf{B}|+|\mathbf{j}|)$, which was slightly modified to

\begin{equation}
E=\sum (|\nabla \cdot \mathbf{B}|+|\mathbf{j}|)^2
\label{ambi:metcalf}
\end{equation}
in an updated version. For computing $j_x$, $j_y$, and $\partial B_z/\partial
z$, a linear force-free model is computed, in the same way as described in
Section~\ref{UHIM}. The method minimizes the functional (\ref{ambi:metcalf})
with the help of a simulated annealing method, which is a robust algorithm to
find a global minimum. In a recent update \citep[published
in][]{metcalf:etal06} the (global) linear force-free assumption has been
relaxed and replaced by local linear force-free assumptions in overlapping
parts of the magnetogram. The method was dubbed {\textit{nonlinear minimum energy
method}, although it does not use true NLFF fields (would be too slow) for
computing the divergence and electric currents. The original linear method
got a fraction of 0.98 of pixels correctly and the nonlinear minimum energy
method even 1.00. Almost all pixels have been correct, except a few on the
boundary (see Figure~\ref{ambi1} fifth row, right panel and last row left
panel, marked with ME1 and ME2, respectively.) Among the fully automatic
methods this approach had the best performance on accuracy. A problem for
practical use of the method was that it is very slow, in particular for the
nonlinear version. Minimum energy methods are routinely used to resolve the
ambiguity in active regions as measured, e.g., with SOT on Hinode or HMI on
SDO.

\subsection{Summary of automatic methods}

The potential-field acute angle method is easy to implement and fast, but its
performance of 0.64\,--\,0.75 is relatively poor. The method is, however, very
important as an initial step for more sophisticated methods. Using more
sophisticated reference fields (linear force-free fields, constant shear,
non-potential fields) in the acute angle method improves the performance to
about 0.83\,--\,0.90. Linear force-free or similar fields are a better
approximation to a suitable reference field, but the corresponding
assumptions are not fulfilled in a strict sense, which prevents a higher
performance. The magnetic pressure gradient and pseudo-current methods are
more difficult to implement as simple acute angle methods, but do not perform
significantly better. A higher performance is prevented, because the basic
assumptions are usually not fulfilled in the entire region. For example, the
assumption that the magnetic pressure always decreases with height is not
fulfilled over bald patches \citep{titov:etal93}. The multi-step U.~Hawai'i
iterative method and the minimum energy methods showed the highest
performance of $>0.97$. The pseudo-current method is in principle similar to
the better performing minimum energy methods, but due to several local minima
it is not guaranteed that the method will always find the global minimum. Let
us remark that \cite{metcalf:etal06} introduced more comparison metrics,
which, however, do not influence the relative rating of the discussed
ambiguity algorithms. They also carried out another test case using the
\cite{chiu:etal77} linear force-free model, for which most of the codes
showed an absolutely better performance, but again this does hardly influence
the relative performance of the different methods. One exception was the
improved non-potential magnetic field algorithm, which performed with similar
excellence as the minimum energy and U.~Hawai'i iterative methods.
Consequently these three methods are all suitable candidates for application
to data. It is, however, not entirely clear to what extent these methods can
be applied to full-disk vector magnetograms and what kind of computer
resources are required.

\subsubsection{Effects of noise and spatial resolution}

The comparison of ambiguity removal methods started in
\cite{metcalf:etal06} has been continued in \cite{leka:etal09}. The
authors investigated the effects of Poisson photon noise and a
limited spatial resolution. It was found that most codes can deal
well with random noise and the ambiguity resolution results are
mainly affected locally, but bad solutions (which are locally wrong
due to noise) do not propagate within the magnetogram. A limited
spatial resolution leads to a loss of information about the fine
structure of the magnetic field and erroneous ambiguity solutions.
Both photon noise and binning to a lower spatial resolution can lead
to artificial vertical currents. The combined effect of noise and
binning affect the computation of a reference magnetic field used in
acute angle methods as well as quantities in minimization approaches
like the electric current density and $\nabla\cdot\mathbf{B}$.
Sophisticated methods based on minimization schemes performed again
best in the comparison of methods and are more suitable to deal with
the additional challenges of noise and limited resolution. As a
consequence of these results \cite{leka:etal09} suggested that one
should use the highest possible resolution for the ambiguity
resolution task and if binning of the data is necessary, this should
be done only after removing the ambiguity. Recently
\cite{georgoulis12} challenged their conclusion that the limited
spatial resolution was the cause of the failure of ambiguity removal
techniques using potential or non-potential reference fields.
\cite{georgoulis12} pointed out that the failure was caused by a
non-realistic test-data set and not by the limited spatial
resolution. This debate has been continued in a reply by
\cite{leka:etal12}. We aim to follow the ongoing debate and provide
an update on this issue in due time.

\subsubsection{HAO AZAM method}

This is an interactive tool, which needs human intervention for the ambiguity
removal. In the test case, which has been implemented and applied by Bruce
Lites, all pixels have been identified correctly. It is of course difficult
to tell about the performance of the method, but only about a human and
software combination. For some individual or a few active regions the method
might be appropriate, but not for a large amount of data.

\subsubsection{Ambiguity removal methods using additional
  observations}

The methods described so far use as input the photospheric magnetic field
vector measured at a single height in the photosphere. If additional
observations/measurements are available they can be used for the ambiguity
removal. Measurements at different heights in order to solve the ambiguity
problem have been proposed by \cite{li:etal93} and revisited by
\cite{li:etal07}. Knowledge of the magnetic field vector at two heights
allows us to compute the divergence of the magnetic field and the method was
dubbed {\textit{divergence-free method}. The method is non-iterative and thus
fast. \cite{li:etal07} applied the method to the same flux-rope simulation by
\cite{fan:etal04} as discussed in the examples above, and the method
recovered about a fraction of 0.98 pixels correctly. The main shortcoming
of this method is certainly that it can be applied only if vector magnetic
field measurements at two heights are available, which is unfortunately not
the case for most current data sets.

\cite{martin:etal08} developed the so-called chirality method for the
ambiguity removal, which takes additional observations into account, e.g.,
H$\alpha$, EUV, or X-ray images. Such images are used to identify the
chirality in solar features like filaments, fibrils, filament channels, or
coronal loops. \cite{martin:etal08} applied the method to different solar
features, but to our knowledge the method has not been tested with synthetic
data, where the true solution of the ambiguity is known. Therefore,
unfortunately one cannot compare the performance of this method with the
algorithms described above. It is also now obvious that fully automatic
feature recognition techniques to identify the chirality from observed images
need to be developed.

After the launch of Solar Orbiter additional vector magnetograms will become
available from above the ecliptic. Taking these observations from two vantage
positions combined is expected to be helpful for the ambiguity resolution. If
separated by a certain angle, the definition of line-of-sight field and
transverse field will be very different from both viewpoints. Removing the
ambiguity should be a straightforward process by applying the transformation
to vertical and horizontal fields on the photosphere from both viewpoints
separately. If the wrong azimuth is chosen, then both solutions will be very
different and the ambiguity can be removed by simply checking the consistency
between vertical and horizontal fields from both observations.

\subsection{Derived quantities, electric currents, and $\alpha$}

The well-known large uncertainties in the horizontal magnetic field
component, in particular in weak field regions  (see Section
\ref{inversion_code}), cause large errors when computing the
electric current density with finite differences via Equation
(\ref{j_photo}). Even more critical is the computation of $\alpha$
with Equation~(\ref{alpha0direct}) in weak field regions and in
particular along polarity inversion lines \citep[see
e.g.,][]{cuperman:etal91}.
The nonlinear force-free coronal magnetic field extrapolation is a boundary
value problem. As we will see later, some of the NLFFF codes make use of
Equation~(\ref{alpha0direct}) to specify the boundary conditions while other
methods use the photospheric magnetic field vector more directly to
extrapolate the field into the corona.

\subsection{Consistency criteria for force-free boundary conditions}
\label{consistency}

After Stokes inversion  (see Section~\ref{inversion_code}) and
azimuth ambiguity removal, we derive the photospheric magnetic field
vector. Unfortunately there might be a problem, when we want to use
these data as the boundary condition for NLFFF extrapolations. Due
to \cite{metcalf:etal95} the solar magnetic field is not force-free
in the photosphere (finite $\beta$ plasma), but becomes force-free
only at about 400~km above the photosphere. This is also visible in
Figure~\ref{figbeta} from \cite{gary01}, which shows the
distribution of the plasma $\beta$ value with height. Consequently
the assumption of a force-free magnetic field is not necessarily
justified in the photosphere. Unless we have information on the
magnetic flux through the lateral and top boundaries, we have to
assume that the photospheric magnetic flux is balanced
\begin{equation}
\int_{S} B_z(x,y,0) \;dx\,dy =0,
\label{flux_balance}
\end{equation}
which is usually the case when
taking an entire active region as the field of view.
In the following we review some necessary conditions the magnetic field
vector has to fulfill in order to be suitable as boundary conditions for
NLFFF extrapolations. \cite{molodensky69,molodensky74} and \cite{aly89}
defined several integral relations, which are related to two moments of the
magnetic stress tensor.
\begin{enumerate}
\item The first moment corresponds to the net magnetic force,
which has to vanish on the boundary:
\begin{eqnarray}
\int_{S} B_x B_z \;dx\,dy = \int_{S} B_y B_z \;dx\,dy = 0 , &&
\label{prepro1} \\
\int_{S} (B_x^2 + B_y^2) \; dx\,dy = \int_{S} B_z^2 \; dx\,dy. &&
\label{prepro2}
\end{eqnarray}
\item
The second moment corresponds to a vanishing torque on the boundary:
\begin{eqnarray}
\int_{S} x \; (B_x^2 + B_y^2) \; dx\,dy &=& \int_{S} x \; B_z^2 \; dx\,dy ,
\label{prepro3} \\
\int_{S} y \; (B_x^2 + B_y^2) \; dx\,dy &=& \int_{S} y \; B_z^2 \; dx\,dy ,
\label{prepro4} \\
\int_{S} y \; B_x B_z \; dx\,dy &=& \int_{S} x \; B_y B_z \; dx\,dy .
\label{prepro5}
\end{eqnarray}
\end{enumerate}
The total energy of a force-free configuration
can be estimated directly from boundary conditions with
the help of the virial theorem \citep[see, e.g.,][for a derivation of
this formula]{aly89}
\begin{equation}
E_{\rm tot}=\frac{1}{\mu_0} \; \int_{S} (x \; B_x + y \; B_y) \; B_z \; dx\,dy .
\label{eq:virial}
\end{equation}
For Equation~(\ref{eq:virial}) to be applicable, the boundary conditions must
be compatible with the force-free assumption. If the integral
relations~(\ref{prepro1})\,--\,(\ref{prepro5}) are not fulfilled then the
data are not consistent with the assumption of a force-free field. A
principal way to avoid this problem would be to measure the magnetic
field vector in the low-$\beta$ chromosphere, but unfortunately such
measurements are not routinely available. We have therefore to rely on
photospheric measurements and apply some procedure, dubbed
`preprocessing', in order to derive suitable boundary conditions for
NLFFF extrapolations.
As pointed out by \cite{aly89} the condition that $\alpha$ is constant on
magnetic field lines (\ref{Bgradalpha}) leads to the integral relation
\begin{equation}
\int_{S_{+}} f(\alpha) \, B_n \cdot dA=\int_{S_{-}} f(\alpha) \, B_n \cdot dA,
\label{flux_balance_alpha}
\end{equation}
where $S_{+}$ and $S_{-}$ correspond to areas with positive and negative
$B_z$ in the photosphere, respectively, and $f$ is an arbitrary function.
Condition (\ref{flux_balance_alpha}) is referred to as differential
flux-balance condition as it generalizes the usual flux-balance
condition~(\ref{flux_balance}). As the connectivity of magnetic field
lines (magnetic positive and negative regions on the boundary
connected by field lines) is a priori unknown,
relation~(\ref{flux_balance_alpha}) is usually only evaluated after a
3D force-free model has been computed.

\subsection{Preprocessing}
\label{preprocessing}

\cite{wiegelmann:etal06} developed a numerical algorithm in order
to use the integral relations~(\ref{prepro1})\,--\,(\ref{prepro5}) to
derive suitable NLFFF boundary conditions from photospheric measurements.
To do so, we define the functional:
\begin{equation}
L_{\mathrm{prep}} = \mu_1 L_1 + \mu_2 L_2 + \mu_3 L_3 + \mu_4 L_4 ,
\end{equation}
\begin{eqnarray}
L_1 &=& \left[
\left(\sum_p B_x B_z \right)^2
+\left(\sum_p B_y B_z \right)^2
+\left(\sum_p B_z^2-B_x^2-B_y^2 \right)^2
\right]  , \\
L_2 &=&
\left[
\left(\sum_p x \left(B_z^2-B_x^2-B_y^2 \right) \right)^2
+\left(\sum_p y \left(B_z^2-B_x^2-B_y^2 \right) \right)^2 \right.
\nonumber \\
& & \left. \hspace*{0.8em}
+\left(\sum_p y B_x B_z -x B_y B_z \right)^2
\right]  , \\
L_3 &=& \left[
\sum_p \left(B_x-B_{x\, \mathrm{obs}} \right)^2
+\sum_p \left(B_y-B_{y\, \mathrm{obs}} \right)^2 \right.
\nonumber \\
&& \left.
+\sum_p \left(B_z-B_{z\, \mathrm{obs}} \right)^2
\right] , \\
L_4 &=& \left[
\sum_p \left(\Delta B_x \right)^2 +\left(\Delta B_y \right)^2 +\left(\Delta B_z \right)^2
\right] .
\end{eqnarray}
The first and second terms ($L_1, L_2$) are quadratic forms of the
force and torque balance conditions, respectively. The $L_3$ term
measures the difference between the measured and preprocessed data.
$L_4$ controls the smoothing, which is useful for the application of
the data to finite-difference numerical code and also because the
chromospheric low-$\beta$ field is smoother than in the photosphere.
The aim is to minimize $L_{\rm prep}$ so that all terms $L_n$ are
made small simultaneously. The optimal parameter sets $\mu_n$ have
to be specified for each instrument separately. The resulting
magnetic field vector is then used to prescribe the boundary
conditions for NLFFF extrapolations. In an alternative approach
\cite{fuhrmann:etal07} applied a simulated annealing method to
minimize the functional. Furthermore they removed the $L_3$ term in
favor of a different smoothing term $L_4$, which uses the median
value in a small window around each pixel for smoothing. The
preprocessing routine has been extended in \cite{wiegelmann:etal08}
by including chromospheric measurements, e.g., by minimizing
additionally the angle between the horizontal magnetic field and
chromospheric H$\alpha$ fibrils. In principle one could add
additional terms to include more direct chromospheric observations,
e.g., line-of-sight measurements of the magnetic field in higher
regions as provided by SOLIS. In principle it should be possible to
combine methods for ambiguity removal and preprocessing in one code,
in particular for ambiguity codes which also minimize a functional
like the \cite{metcalf94} minimum energy method. A mathematical
difficulty for such a combination is, however, that the
preprocessing routines use continuous values, but the ambiguity
algorithms use only two discrete states at each pixel.
Preprocessing minimizes the integral relations
(\ref{prepro1}\,--\,\ref{prepro5}) and the value of these integrals
reduces usually by orders of magnitudes during the preprocessing
procedure. These integral relation are, however, only necessary and
not sufficient conditions for force-free consistent boundary
conditions, and preprocessing does not make use of condition
(\ref{flux_balance_alpha}). Including this condition is not straight
forward as one needs to know the magnetic field line connectivity,
which is only available after the force-free configuration has been
computed in 3D. An alternative approach for deriving force-free
consistent boundary conditions is to allow changes of the boundary
values (in particular the horizontal field) during the force-free
reconstruction itself, e.g., as recently employed by
 \cite{wheatland:etal09}, \cite{amari:etal10}, and \cite{wiegelmann:etal10a}.
The numerical implementation of these approaches does necessarily depend
on the corresponding force-free extrapolation codes and we refer to
Sections~\ref{grad_rubin} and \ref{optimization} for details.

\section{Nonlinear Force-free Fields in 3D}

In the following section we briefly discuss some
general properties of force-free fields, which are relevant for
solar physics, like the magnetic helicity, estimations of the
minimum and maximum energy a force-free field can have for certain
boundary conditions and investigations of the stability. Such
properties are assumed to play an important role for solar eruptions.
The Sun and the solar corona are of course three-dimensional and for
any application to observed data, configurations based on symmetry assumptions
(as used in Section~\ref{NLFFF-2D}) are usually not applicable.
The numerical treatment of nonlinear problems, in particular in 3D,
is significantly more difficult than linear ones. Linearized equations
are often an over-simplification which does not allow the appropriate
treatment of physical phenomena. This is also true for force-free
coronal magnetic fields and has been demonstrated by comparing
linear force-free configurations (including potential fields,
where the linear force-free parameter $\alpha$ is zero).

Computations of the photospheric $\alpha$ distribution from measured
vector magnetograms by Equation~(\ref{alpha0direct}) show that
$\alpha$ is a function of space
\citep[see, e.g.,][]{pevtsov:etal94,regnier:etal02,derosa:etal09}.
Complementary to
this direct observational evidence that nonlinear effects are important,
there are also theoretical arguments. Linear models are too simple to
estimate the free magnetic energy. Potential fields correspond to the
minimum energy configuration for a given magnetic flux distribution on
the boundary. Linear force-free fields contain an unbounded magnetic
energy in an open half-space above the photosphere \citep{seehafer78},
because the governing equation in this case is the Helmholtz (wave)
equation (Equation~(\ref{eq:Helmholtz})) whose solution decays slowly
toward infinity. Consequently both approaches are not suitable for the
estimation of the magnetic energy, in particular not an estimation of the
free energy a configuration has in excess of a potential field.

\subsection{Magnetic helicity}
\label{helicity}

Magnetic helicity is a quantity closely related to a property of the
force-free field \citep{woltjer58}, and is defined by
\begin{equation}
H_{\mathrm{m}} = \int_V \mathbf{A}\cdot \mathbf{B}\; dV,
\label{eq:helicity}
\end{equation}
where $\mathbf{B} = \nabla \times \mathbf{A}$ and $\mathbf{A}$ is the vector
potential. When $\mathbf{B}$ is given, $\mathbf{A}$ is not unique and a
gradient of any scalar function can be added without changing
$\mathbf{B}$. Such gauge freedom does not affect the value of
$H_{\mathrm{m}}$ if the volume $V$ is bounded by a magnetic
surface (i.e., no field lines go through the surface).
Figure~\ref{fig:helicity} shows simple torus configurations
and their magnetic helicities. As can be guessed from the figures,
magnetic helicity is a topological quantity describing how the field
lines are twisted or mutually linked, and is conserved when resistive
diffusion of magnetic field is negligible. In the case of the solar
corona, the bottom boundary (the photosphere) is not a magnetic
surface, and field lines go through it. Even under such conditions,
an alternative form for the magnetic helicity which does not depend
on the gauge of $\mathbf{A}$ can be defined \citep{berger-field84,finn-antonsen85}.
On the Sun one finds the hemispheric helicity sign rule
\citep[see, e.g.][and references therein]{pevtsov:etal95,wang:etal10}.
For various features like active regions, filaments, coronal loops
and interplanetary magnetic clouds the helicity is
negative in the northern and positive in the southern
hemisphere.

\epubtkImage{helicity4}{%
\begin{figure}[htbp]
\centerline{\includegraphics[scale=0.7]{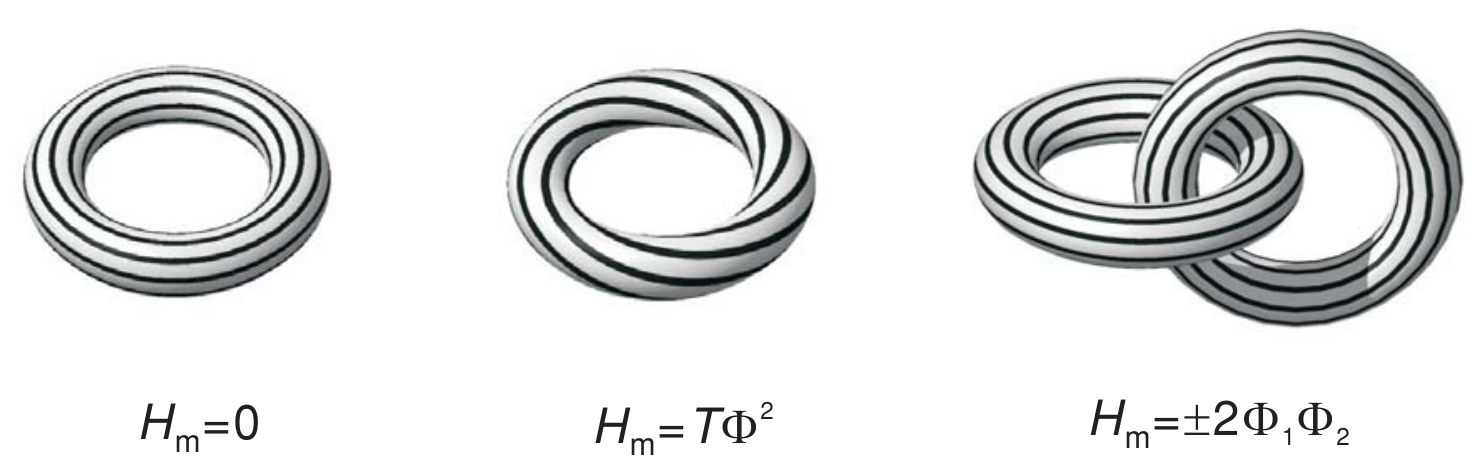}}
\caption{Magnetic helicity of field lines in torus configuration: untwisted
(left), twisted by $T$ turns (middle), and two untwisted but intersecting
tori (right). $\Phi$ stands for the total magnetic flux.}
\label{fig:helicity}
\end{figure}}

\subsection{Energy principles}

Energy principles leading to various magnetic fields (potential fields,
linear force-free fields, and nonlinear force-free fields) were summarized in
\cite{sakurai89}. For a given distribution of magnetic flux ($B_z$) on the
boundary,
\begin{itemize}
\item[(a)] a potential field is the state of minimum energy.
\item[(b)] If the magnetic energy is minimized with an additional condition
of a fixed value of $H_{\mathrm{m}}$, one obtains a linear force-free field.
The value of constant $\alpha$ should be an implicit function of $H_{\mathrm{m}}$.
The obtained solution may or may not be a minimum of energy; in the latter
case the solution is dynamically unstable.
\item[(c)] If the magnetic energy is minimized by specifying the
connectivity of all the field lines, one obtains a nonlinear force-free
field. The solution may or may not be dynamically stable.
\end{itemize}
Item (c) is more explicitly shown by introducing the so-called Euler
potentials $(u, v)$ for the magnetic field \citep{stern70},
\begin{equation}
\mathbf{B} = \nabla u \times \nabla v .
\label{Euler-potential}
\end{equation}
This representation satisfies $\nabla\cdot \mathbf{B} =0$. Since
$\mathbf{B}\cdot\nabla u = \mathbf{B}\cdot\nabla v =0$, $u$ and
$v$ are constant along the field line. The values of $u$ and
$v$ on the boundary can be set so that $B_z$ matches the given
boundary condition. If the magnetic energy is minimized with
the values of $u$ and $v$ specified on the boundary, one
obtains Equation~(\ref{eq:force-free}) for a general
(nonlinear) force-free field.

By the construction of the energy principles, the energy of (b) or
(c) is always larger than that of the potential field (a). If the
values of $u$ and $v$ are so chosen (there is enough freedom) that
the value of $H_{\mathrm{m}}$ is the same in cases (b) and (c), then the
energy of nonlinear force-free fields (c) is larger than that of the
linear force-free field (b). Therefore, we have seen that magnetic
energy increases as one goes from a potential field to a linear
force-free field, and further to a nonlinear force-free field.
Suppose there are field lines with enhanced values of $\alpha$
(carrying electric currents stronger than the surroundings).
By some instability (or magnetic reconnection), the excess energy
may be released and the twist in this part of the volume may diminish.
However, in such rapid energy release processes, the magnetic helicity
over the whole volume tends to be conserved \citep{berger84}. Namely
local twists represented by spatially-varying $\alpha$ only propagate
out from the region and are homogenized, but do not disappear.
Because of energy principle (b), the end state of such relaxation
will be a linear force-free field. This theory
\citep[Taylor relaxation;][]{taylor74,taylor86} explains the
commonly-observed behavior of laboratory plasmas to relax
toward linear force-free fields.
On the Sun this behaviour is not observed, however.
A possible explanation could be that
since we observe spatially-varying
$\alpha$ on the Sun, relaxation to linear force-free fields only
takes place at limited occasions (e.g., in a flare) and over a
limited volume which magnetic reconnection (or other processes) can
propagate and homogenize the twist.

\subsection{Maximum energy}
\label{aly-sturrock}

There is in particular a large interest on force-free configurations for a
given vertical magnetic field $B_n$ on the lower boundary and in which range
the energy content of these configurations can be. For such theoretical
investigations, one usually assumes a so-called star-shaped volume, like the
exterior of a spherical shell and the coronal magnetic field is unbounded but
has a finite magnetic energy. (Numerical computations, on the other hand, are
mainly carried out in finite computational volumes, like a 3D-box in
Cartesian geometry.) It is not the aim of this review to follow the involved
mathematical derivation, which the interested reader finds in \cite{aly84}.
As we saw above, the minimum energy state is reached for a potential field.
On the other hand, one is also interested in the maximum energy a force-free
configuration can obtain for the same boundary conditions $B_n$. This problem
has been addressed in the so-called Aly--Sturrock conjecture
\citep{aly84,aly91,sturrock91}. The conjecture says that the maximum magnetic
energy is obtained if all magnetic field lines are open (have one footpoint
in the lower boundary and reach to infinity). This result implies that any
non-open force-free equilibrium (which contains electric currents parallel to
closed magnetic field lines, e.g., created by stressing closed potential field
lines) contains an energy which is higher than the potential field, but lower
than the open field. As pointed out by \cite{aly91} these results imply that
the maximum energy which can be released from an active region, say in a
flare or coronal mass ejection (CME), is the difference between the energy of
an open field and a potential field. While a flare requires free magnetic
energy, the Aly--Sturrock conjecture does also have the consequence that it
would be impossible that all field lines become open directly after a flare,
because opening the field lines costs energy. This is in some way a
contradiction to observations of CMEs, where a closed magnetic structure
opens during the eruption. \cite{choe:etal02} constructed force-free
equilibria containing tangential discontinuities in multiple flux systems,
which can be generated by footpoint motions from an initial potential field.
These configurations contain energy exceeding the open field, a violation of
the Aly--Sturrock conjecture, and would release energy by opening all field
lines. Due to the tangential discontinuities, these configurations contain
thin current sheets, which can develop micro-instabilities to convert
magnetic energy into other energy forms (kinetic and thermal energy) by
resistive processes like magnetic reconnection.
It is not clear \citep{aly07}, however, which conditions are exactly necessary
to derive force-free fields with
energies above the open field: Is it necessary that the multiple flux-tubes
are separated by non-magnetic regions like in \cite{choe:etal02}? Or would it
be sufficient that the field in this region is much weaker than in the flux
tubes but remains finite? \citep[See][for a related discussion.]{sakurai07}

\subsection{Stability of force-free fields}

In principle the MHD stability criteria can also be applied to force-free
equilibria. Typical approaches \citep[see the book by][]{priest82} to
investigate the stability of
ideal MHD equilibria (which correspond to the assumption
of infinite electrical conductivity)
are normal mode analysis and an
energy criterion. The basic question is how a small disturbance to the
equilibrium evolves. Analytic methods typically linearize the problem around
an equilibrium state, which leads to the so-called linear stability analysis.
One has to keep in mind, however, that a linearly-stable configuration might
well be nonlinearly unstable. The nonlinear stability of a system is usually
investigated numerically with the help of time dependent simulations, e.g.,
with an MHD code (see also Section~\ref{numerical_stability} for an
application to NLFFF equilibria). In the following we concentrate on linear
stability investigations by using an energy criterion.

For a force-free configuration the energy is given by
\begin{equation}
W_0=\int \frac{B_0^2}{2 \mu_0} dV
\label{stability_w0}
\end{equation}
where the subscript $0$ corresponds to the equilibrium state.
This equilibrium becomes disturbed by an displacement
$\boldsymbol{\xi}(r_0,t)$ in the form $ \mathbf{B}=\mathbf{B}_0+\mathbf{B}_1$
with $\mathbf{B}_1=\nabla_0 \times (\boldsymbol{\xi} \times \mathbf{B}_0)$.
This form of the magnetic field displacement has its origin
from the linearized induction equation
$\frac{\partial \mathbf{B}_1}{\partial t}=\nabla \times (\mathbf{v}_1 \times \mathbf{B}_0)$,
where the velocity field has been replaced by the displacement $\boldsymbol{\xi}$.
The MHD energy principle \citep{bernstein:etal58} reduces for force-free
fields to \citep{molodensky74}:
\begin{equation}
W=\frac{1}{2\mu_0} \, \int_V
\left[
\left(\nabla \times (\boldsymbol{\xi} \times \mathbf{B}) \right)^2
-\left(\nabla \times (\boldsymbol{\xi} \times \mathbf{B}) \right) \cdot
\left(\boldsymbol{\xi} \times (\nabla\times \mathbf{B})\right)
\right] \; dV.
\label{stable:energy1}
\end{equation}
A configuration is stable if $W>0$, unstable for $W<0$ and marginally
stable for $W=0$. For force-free fields and using the perturbed
vector potential $\mathbf{A}_1=\boldsymbol{\xi} \times \mathbf{B}$,
Equation~(\ref{stable:energy1}) can be written as:
\begin{equation}
W=\frac{1}{2\mu_0} \, \int_V
\left[ \left(
\nabla \times \mathbf{A}_1 \right)^2 -\alpha \mathbf{A}_1 \cdot \nabla \times \mathbf{A}_1
\right] \; dV .
\label{stable:energy2}
\end{equation}
From Equation~(\ref{stable:energy2}) it is obvious that the potential
field with $\alpha=0$ is stable. If we approximate
$|\nabla \times \mathbf{A}_1| \sim |A_1|/\ell$ with a typical
length scale $\ell$ of the system, the first term may remain
larger than the second term (i.e., stability) in
Equation~(\ref{stable:energy2}) if
\begin{equation}
|\alpha|
\,\mbox{\raisebox{0.3ex}{$<$}\hspace{-0.8em}\raisebox{-0.7ex}{$\sim$}}\,
1/\ell .
\end{equation}
This means that the scale of twist in the system, $1/\alpha$,
should be larger than the system size $\ell$ for it to be stable.
This criterion is known as Shafranov's limit in plasma physics.
More precise criteria for stability can be obtained for specific
geometries. For example the case of cylindrical linear force-free
field (Lundquist's field) was studied by \citet{goedbloed-hagebeuk72}.

\epubtkImage{tdtoeroek}{%
\begin{figure}[htbp]
\centerline{\includegraphics[width=12cm]{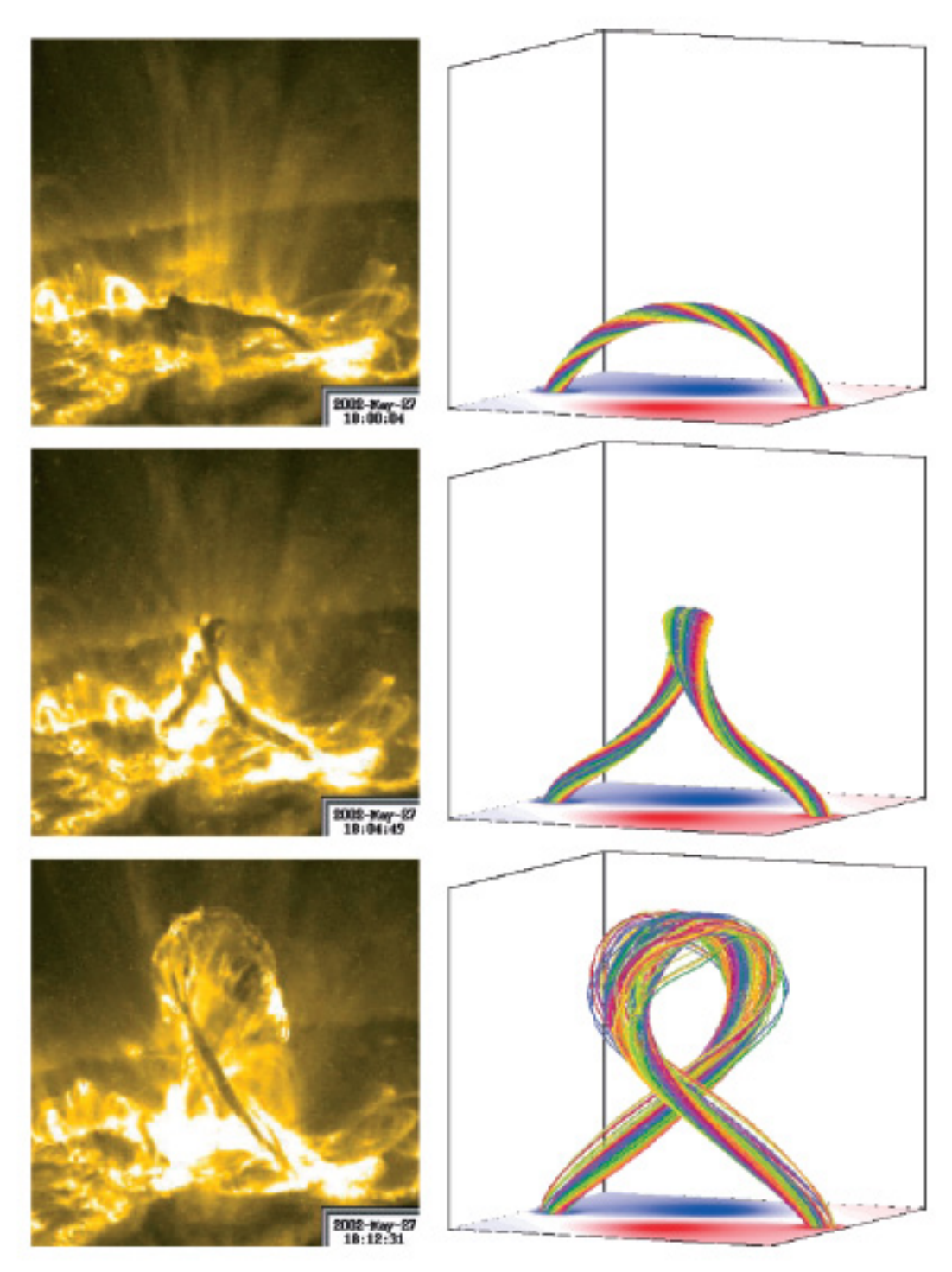}}
\caption{Numerical simulations starting from an unstable branch of the
  Titov--D\'emoulin equilibrium in comparison with TRACE observations
  of an eruption. The original figure was published as Figure~1 in
  \cite{toeroek:etal05}.}
\label{td3}
\end{figure}}

\subsection{Numerical stability investigations}
\label{numerical_stability}

\cite{toeroek:etal05} investigated the stability of the nonlinear
force-free Titov--D\'emoulin equilibrium numerically with the help of
a time-dependent MHD code. Figure~\ref{td3} shows snapshots from
MHD simulation starting from in an unstable branch of the
Titov--D\'emoulin equilibrium in comparison with a solar eruption
observed with TRACE. The simulation shows a very good agreement
with the observed eruptions and indicates that a helical kink
instability can trigger coronal eruptions. Dependent on particular
parameters of the original Titov--D\'emoulin equilibrium the eruption
remains confined or leads to a coronal mass ejection
\citep[see][for details]{toeroek:etal05}.

%
%
\newpage
\section{Numerical Methods for Nonlinear Force-free Fields}

In the following we review five different approaches for the
computation of nonlinear force-free coronal magnetic fields.
The aim of all codes is to extrapolate
photospheric vector field measurements into the corona, but
the way how the measurements are used is different.
MHD relaxation and optimization methods prescribe the
three components of the magnetic field vector on the bottom
boundary. Grad--Rubin methods use the
vertical magnetic field and the vertical electric current
density (or $\alpha$-distribution) as boundary condition.
The upward integration method and the boundary-element method
require a combination of both boundary conditions.
In the following we will briefly
discuss the main features of these five methods. Grad--Rubin,
MHD relaxation, and optimization methods require first the
computation of a potential field; then the appropriate boundary
conditions are specified and eventually one iterates numerically
for a solution of the NLFFF equations. Upward integration and
boundary-element methods do not require first the computation of
a potential field, but solve the NLFFF equations more directly. Both
methods have, however, some shortcomings as explained later.
Often one is interested anyway to get also the potential field,
e.g., to derive the energy the NLFFF field has in excess of the potential
field. A more detailed review on the mathematical and computational
implementations and recent code updates can be found in \cite{wiegelmann08}.

\subsection{Upward integration method}

This straightforward method was proposed by \cite{nakagawa74} and
it has been first computationally implemented by \cite{wu:etal85,wu:etal90}.
The basic idea of this method is to reformulate
Equations~(\ref{forcebal})\,--\,(\ref{solenoidal}) and extrapolate the
magnetic field vector into the solar corona. The method
is not iterative and extrapolates the magnetic field directly
upward, starting from the bottom layer, where the field is measured.
From $\mathbf{B}_0(x,y,0)$ one computes the $z$-component of the
electric current $\mu_0 j_{z0}$ by Equation~(\ref{j_photo}) and
the corresponding $\alpha$-distribution with Equation
(\ref{alpha0direct}). Then the $x$ and $y$-components of
the electric current are calculated by Equation~(\ref{amperealpha}):
\begin{eqnarray}
\mu_0 j_{x0 } & = & \alpha_0\, B_{x0} \,,\\
\mu_0 j_{y0} & = & \alpha_0\, B_{y0} \,.
\end{eqnarray}
Finally, we get the $z$-derivatives of the magnetic field vector with
Equations~(\ref{ampere}) and (\ref{solenoidal}) as
\begin{eqnarray}
\frac{\partial B_{x0}}{\partial z} & = &
\mu_0 j_{y0}+ \frac{\partial B_{z0}}{\partial x} \,, \\
\frac{\partial B_{y0}}{\partial z} & = &
\frac{\partial B_{z0}}{\partial y}- \mu_0 j_{x0} \,,\\
\frac{\partial B_{z0}}{\partial z} & = &
-\frac{\partial B_{x0}}{\partial x}- \frac{\partial B_{y0}}{\partial y} \,.
\end{eqnarray}
A numerical integration provides the magnetic field vector at the
level $z+dz$. These steps are repeated in order to integrate the
equations upwards in $z$. Naively one would assume to derive finally
the 3D magnetic fields in the corona, which is indeed the idea of
this method. The main problem is that this simple straightforward
approach does not work because the method is mathematically ill-posed
and the algorithm is unstable (see, e.g., \citealp{cuperman:etal90}
and \citealp{amari:etal97} for details). As a result of this numerical
instability one finds an exponential growth of the magnetic field w
ith increasing height. The reason for this is that the method
transports information only from the photosphere upwards. Other
boundary conditions, e.g., at an upper boundary, either at a finite
height or at infinity cannot be taken into account. Several attempts
have been made to stabilize the algorithm, e.g., by smoothing and
reformulating the problem with smooth analytic functions
\citep[e.g.,][]{cuperman:etal91,demoulin:etal92,song:etal06}.
Smoothing does help somewhat to diminish the effect of growing
modes, because the shortest spatial modes are the fastest growing
ones. To our knowledge the upward integration method has not been
compared in detail with other NLFFF codes and it is therefore hard
to evaluate the performance of this method.

\subsection{Grad--Rubin method}
\label{grad_rubin}

The Grad--Rubin method has been originally proposed (but not numerically
implemented) by \cite{grad:etal58} for the application to
fusion plasma. The first numerical application to coronal magnetic
fields was carried out by \cite{sakurai81}. The original Grad--Rubin
approach uses the $\alpha$-distribution on one polarity and the initial
potential magnetic field to calculate the electric current density with
Equation~(\ref{Bgradalpha}) and to update the new magnetic field $\mathbf{B}$
from the Biot--Savart equation (\ref{amperealpha}). This scheme is repeated
iteratively until a stationary state is reached, where the magnetic field
does not change anymore. \cite{amari:etal97,amari:etal99} implemented the
Grad--Rubin method on a finite difference grid and decomposes
Equations~(\ref{forcebal})\,--\,(\ref{solenoidal}) into a hyperbolic
part for evolving $\alpha$ along the magnetic field lines and an
elliptic one to update the magnetic field from Ampere's law:
\begin{eqnarray}
\mathbf{B}^{(k)} \cdot \nabla \alpha^{(k)} &= &0 \,, \\
\alpha^{(k)}|_{S^{\pm}} &=& \alpha_{0\pm} \,.
\end{eqnarray}
This evolves $\alpha$ from one polarity on the boundary along
the magnetic field lines into the volume above. The value of
$\alpha_{0\pm}$ is given either in the positive or negative polarity:
\begin{eqnarray}
\nabla \times \mathbf{B}^{(k+1)} &=& \alpha^{(k)} \mathbf{B}^{(k)}  \\
\nabla \cdot \mathbf{B}^{(k+1)} &=& 0 \,, \\
B^{(k+1)}_z|_{S^{\pm}} &=& B_{z0} \,, \\
{\rm \lim_{|r|\rightarrow\infty}}|\mathbf{B}^{(k+1)}| &=& 0 \,.
\end{eqnarray}
An advantage from a mathematical point of view is that the Grad--Rubin
approach solves the nonlinear force-free equations as a well-posed
boundary value problem.
As shown by \cite{bineau72} the Grad--Rubin-type
boundary conditions, the vertical magnetic field and for one polarity the
distribution of $\alpha$, ensure the existence and unique NLFFF solutions
at least for small values of $\alpha$ and weak nonlinearities. See
\cite{amari:etal97,amari:etal06} for more details on the mathematical
aspect of this approach. The largest-allowed current and the corresponding
maximum values of $\alpha$ for which one can expect convergence of the
Grad--Rubin approach have been studied in \cite{inhester:etal06}.
Starting from an initial potential field the NLFFF equations are
solved iteratively in the form of
Equations~(\ref{amperealpha})\,--\,(\ref{Bgradalpha}). The horizontal
component of the measured magnetic field is then used to compute the
distribution of $\alpha$ on the boundary using
Equation~(\ref{alpha0direct}). While $\alpha$ is computed this way on
the entire lower boundary, the Grad--Rubin method requires only the
prescription of $\alpha$ for one polarity. For measured data which
contain noise, measurement errors, finite forces, and other
inconsistencies the two solutions can be different: However, see for
example the extrapolations from Hinode data carried out in
\cite{derosa:etal09}. While both solutions are based on well-posed
mathematical problems, they are not necessary consistent with the
observations on the entire lower boundary. One can check the
consistency of the $\alpha$-distribution on both polarities with
Equation~(\ref{flux_balance_alpha}).

As a further step to derive one unique solution the Grad--Rubin approach
has been extended by \cite{wheatland:etal09} and \cite{amari:etal10} by
using these two different solutions (from different polarities) to correct
the $\alpha$-distribution on the boundaries and to find finally one
consistent solution by an outer iterative loop, which changes the
$\alpha$-distribution on the boundary. An advantage in this approach
is that one can specify where the $\alpha$-distribution, as computed
by Equation~(\ref{alpha0direct}), is trustworthy (usually in strong
field regions with a low measurement error in the transverse field)
and where not (in weak field regions). This outer iterative loop,
which aims at finding a consistent distribution of $\alpha$ on both
polarities, allows also to specify where the initial distribution
of $\alpha$ is trustworthy.

\subsection{MHD relaxation method}

MHD relaxation method means that a reduced set of time-dependent MHD
equations is used to compute stationary equilibria:
\begin{eqnarray}
\nu \mathbf{v}&=& (\nabla \times \mathbf{B}) \times \mathbf{B} , \label{eqmotion}\\
\mathbf{e} + \mathbf{v} \times \mathbf{B} &=& \mathbf{0} , \label{iohms}\\
\frac{\partial \mathbf{B}}{\partial t} &=&-\nabla \times \mathbf{e} ,
\label{faraday} \\
\nabla \cdot \mathbf{B} &=& 0
\label{sole2}.
\end{eqnarray}
Here $\nu$ is a fictitious viscosity,
$\mathbf{v}$ the fluid velocity and $\mathbf{e}$ the electric field.
For general MHD equilibria the approach was proposed by \cite{chodura:etal81}.
Applications to force-free coronal magnetic fields can be found in
\cite{mikic:etal94}, \cite{roumeliotis96}, and \cite{mcclymont:etal97}.
In principle any time-dependent MHD code can be used for this aim.
The first NLFFF implementation of this methods used the code developed
by \cite{mikic:etal88}. MHD relaxation means that an initial
non-equilibrium state is relaxed towards a stationary state, here NLFFF.
The initial non-equilibrium state is often a potential field in the
3D-box, where the bottom boundary field has been replaced by the measurements.
This leads to large deviations from the equilibrium close to this boundary.
As a consequence one finds a finite
plasma flow velocity $\mathbf{v}$ in Equation~(\ref{eqmotion})
because all
non-magnetic forces accumulate in the velocity field.
This velocity field is reduced during the
relaxation process and the
force-free field equations are obviously fulfilled when
the left-hand side of Equation~(\ref{eqmotion}) vanishes.
The viscosity $\nu$ is usually chosen as
\begin{equation}
\nu = \frac{1}{\mu} \, |\mathbf{B}|^2
\label{nudef}
\end{equation}
with $\mu=$ constant. By combining Equations~(\ref{eqmotion}), (\ref{iohms}),
(\ref{faraday}), and (\ref{nudef}) one gets a relaxation process for
the magnetic field
\begin{equation}
\frac{\partial \mathbf{B}}{\partial t} =\mu \; \mathbf{F}_{\rm MHD} ,
\label{relaxinduct}
\end{equation}
\begin{equation}
\mathbf{F}_{\rm MHD} =\nabla \times \left(\frac{\left[(\nabla \times \mathbf{B})
\times \mathbf{B}\right] \times \mathbf{B}}{B^2} \right).
\label{def_fmhd}
\end{equation}
For details regarding a currently-used implementation of
this approach see \cite{valori:etal05}.

\subsection{Optimization approach}
\label{optimization}

The optimization approach as proposed in \cite{wheatland:etal00} is
closely related to the MHD relaxation approach. It shares with
this method that a similar initial non-equilibrium state is iterated
towards a NLFFF equilibrium. It solves a similar iterative equation
as Equation~(\ref{relaxinduct})
\begin{equation}
\frac{\partial \mathbf{B}}{\partial t} = \mu \; \mathbf{F},
\end{equation}
but $\mathbf{F}$ has additional terms, as explained below. The force-free
and solenoidal conditions are solved by minimizing the functional
\begin{equation}
L=\int_{V} \left[B^{-2} \, |(\nabla \times \mathbf{B}) \times \mathbf{B}|^2 +|\nabla \cdot \mathbf{B}|^2\right] \; dV.
\label{defL}
\end{equation}
If the minimum of this functional at $L=0$ is attained then the NLFFF
equations (\ref{forcebal})\,--\,(\ref{solenoidal}) are fulfilled.
The functional is minimized by taking the functional derivative
of Equation~(\ref{defL}) with respect to an iteration parameter $t$:
\begin{equation}
\frac{1}{2} \; \frac{d L}{d t}=-
\int_{V} \frac{\partial \mathbf{B}}{\partial t} \cdot \mathbf{F} \; dV
-\int_{S} \frac{\partial \mathbf{B}}{\partial t} \cdot \mathbf{G} \; dS ,
\label{minimize1}
\end{equation}
\begin{eqnarray}
\mathbf{F} & = &
\nabla \times \left(
\frac{\left[(\nabla \times \mathbf{B}) \times \mathbf{B} \right] \times \mathbf{B}}{B^2} \right)
\nonumber\\
& & + \; \left\{
-\nabla \times \left(\frac{((\nabla \cdot \mathbf{B}) \; \mathbf{B}) \times \mathbf{B}}{B^2} \right)
\right. \nonumber\\
& & - \mathbf{\Omega} \times (\nabla \times \mathbf{B}) -\nabla(\mathbf{\Omega} \cdot \mathbf{B})
\nonumber\\
& & + \left. \mathbf{\Omega}(\nabla \cdot \mathbf{B}) + \Omega^2 \; \mathbf{B} \right\}
,
\label{wheateq}
\end{eqnarray}
\begin{equation}
\mathbf{\Omega} =
B^{-2} \;\left[
(\nabla \times \mathbf{B}) \times \mathbf{B}-(\nabla \cdot \mathbf{B}) \; \mathbf{B}
\right] .
\end{equation}
For vanishing surface terms the functional $L$ decreases monotonically if
the magnetic field is iterated by
\begin{equation}
\frac{\partial \mathbf{B}}{\partial t} = \mu \; \mathbf{F}.
\end{equation}
The first term in Equation~(\ref{wheateq}) is identical
with $\mathbf{F}_{\rm MHS}$ as defined in Equation~(\ref{def_fmhd}).

A principal problem with the optimization and the MHD-relaxation
approaches is that using the full magnetic field vector on the lower
boundary does not guarantee the existence of a force-free configuration
(see the consistency criteria in Section~\ref{consistency}).
Consequently, if fed with inconsistent boundary data, the codes
cannot find a force-free configuration, but a finite residual Lorentz
force and/or a finite divergence of the field remains in the 3D equilibrium.
A way around this problem is to preprocess the measured photospheric data,
as explained in Section~\ref{preprocessing}. An alternative approach is
that one allows deviations of the measured horizontal field vector and
the corresponding field vector on the lower boundary of the computational
box during the minimization of the functional (\ref{defL}).
\cite{wiegelmann:etal10a} extended this functional by another term
\begin{equation}
\nu \int_{S} (\mathbf{B} - \mathbf{B}_{\rm obs})\cdot\mathbf{W}\cdot(\mathbf{B} - \mathbf{B}_{\rm obs})\; dS
\end{equation}
where $\nu$ is a free parameter and the matrix $\mathbf{W}$ contains information
how reliable the data (mainly measurements of the horizontal photospheric
field) are. With this approach inconsistencies in the measurement lead to
a solution compatible with physical requirements (vanishing Lorentz force
and divergence), leaving differences between $\mathbf{B}_{\rm obs}$ and the
bottom boundary field $\mathbf{B}$ in regions where $\mathbf{W}$ is low (and the
measurement error high). Consequently this approach takes measurement errors,
missing data, and data inconsistencies into account. Further tests are
necessary to investigate whether this approach or preprocessing, or a
combination of both, is the most effective way to deal with noisy and
inconsistent photospheric field measurements. This approach, as well
as a variant of the Grad--Rubin method, have been developed in response
to a joint study by \cite{derosa:etal09}, where one of the main findings w
as that force-free extrapolation codes should be able to incorporate
measurement inconsistencies (see also Section~\ref{nlfff_consortium}).

\subsection{Boundary-element methods}

The boundary-element method was developed by \cite{yan:etal00}
and requires the magnetic field vector and the $\alpha$-distribution
on the boundary as input. The NLFFF equations relate the magnetic field
values on the boundary with those in the volume:
\begin{equation}
c_i \mathbf{B}_i= \oint_S
\left(
\mathbf{\bar{Y}} \frac{\partial \mathbf{B}}{\partial n}
- \frac{\partial \mathbf{\bar{Y}}}{\partial n} \mathbf{B}_0
\right) \; dS
\label{yan1}
\end{equation}
with $c_i=1$ for points in the volume and $c_i=1/2$ for boundary points
and $\mathbf{B}_0$ is the magnetic field vector on the boundary, where
\begin{equation}
\mathbf{\bar{Y}}={\rm diag} \left(
\frac{\cos(\lambda_x r)}{4 \pi r},
\frac{\cos(\lambda_y r)}{4 \pi r},
\frac{\cos(\lambda_z r)}{4 \pi r}
\right)
\end{equation}
and $\lambda_i \, (i=x,y,z)$ are implicitly computed with integrals
over the 3D volume,
\begin{equation}
\int_V
Y_i[\lambda_i^2 B_i-\alpha^2 B_i -(\nabla \alpha \times \mathbf{B}_i)] dV=0 .
\label{eq41}
\end{equation}
The boundary-element method is slow for computing the NLFFF in a
3D domain. \cite{rudenko-myshyakov09} raised questions on this method.

\subsection{Comparison of methods and the NLFFF consortium}
\label{nlfff_consortium}

Since 2004 a group of scientists chaired by Karel Schrijver compare,
evaluate, and improve methods for the nonlinear force-free computation
of coronal magnetic fields and related topics.
The test
cases are available at
http://www.lmsal.com/$\sim$derosa/for\_nlfff/
So far six workshops have been organized and the consortium published
four joint publications:
\begin{enumerate}
\item \cite{schrijver:etal06} performed blind tests on analytical
force-free field models with various boundary conditions to show
that in general the NLFFF algorithms perform best where the magnetic
field and the electrical currents are strongest, but they are also
very sensitive to the specified boundary conditions. Nevertheless,
it was shown that the optimization method as proposed by
\cite{wheatland:etal00} and as implemented by \cite{wiegelmann04}
was the fastest-converging and best-performing one for this analytical test case.
\item \cite{metcalf:etal08} tested the performance of the NLFFF algorithms
applied to a solar-like reference model including realistic photospheric
Lorentz forces and a complex magnetic field structure. All the codes
were able to recover the presence of a weakly twisted, helical flux rope.
Due to the sensitivity to the numerical details, however, they were less
accurate in reproducing the field connectivity and magnetic energy when
applied to the preprocessed, more force-free, chromospheric-like boundary
conditions. When applied to the forced, not preprocessed photospheric data
the codes did not perform successfully, indicating that the consistency of
the used boundary conditions is crucial for the success of the magnetic field
extrapolations. It also showed that the magnetic field connection between the
photosphere, chromosphere, and lower corona needs to be additionally
precisely modeled.
\item \cite{schrijver:etal08} used four different codes and a variety of
boundary conditions to compute 14 NLFFF models based on Hinode/SOT-SP%
\epubtkFootnote{Solar Optical Telescope Spectro-Polarimeter}
data of an active region around the time of a powerful flare. When applied
to this real solar data, the models produced a wide variety of magnetic
field geometries, energy contents, and force-freeness.
Force-free consistency criteria, like the
alignment of electric currents with magnetic field lines, have been
best fulfilled for computations with the Grad--Rubin approach.
It was concluded
that strong electrical currents in the form of an ensemble of thin strands
emerge together with magnetic flux preceding the flare. The global patterns
of magnetic fields are compatible with a large-scale twisted flux rope
topology, and they carry energy which is large enough to power the flare
and its associated CME.
\item
\cite{derosa:etal09} found that various NLFFF models differ remarkably
in the field line configuration and produce different estimates of the
free magnetic energy when applied to Hinode/SOT-SP data. This problem
was recognized already in the first application to Hinode data in
\cite{schrijver:etal08} and it has been worked out that a small
field-of-view vector magnetogram, which does not contain an entire
active region and its surroundings, does not provide the necessary
magnetic connectivity for successful NLFFF extrapolations. As visible
in Figure~\ref{nlfff4f1} the stereoscopically-reconstructed loops by
\cite{aschwanden:etal08a} do not agree well with the NLFFF models.
Unfortunately, the FOV of Hinode covered only a small fraction
(about 10\%) of area spanned by loops reconstructed from
STEREO/SECCHI images. The quantitative comparison was unsatisfactory
and NLFFF models have not been better as potential fields here. In
other studies NLFFF methods have shown to be superior to potential
and linear force-free extrapolations \citep{wiegelmann:etal05}. NLFF
field lines showed in particular excellent agreement with the observed
loops, when both footpoints are within the FOV of the vector magnetogram
and sufficiently far away from the boundaries.
\end{enumerate}

\epubtkImage{nlfff4f1}{%
\begin{figure}[htb]
\centerline{\includegraphics[width=\textwidth]{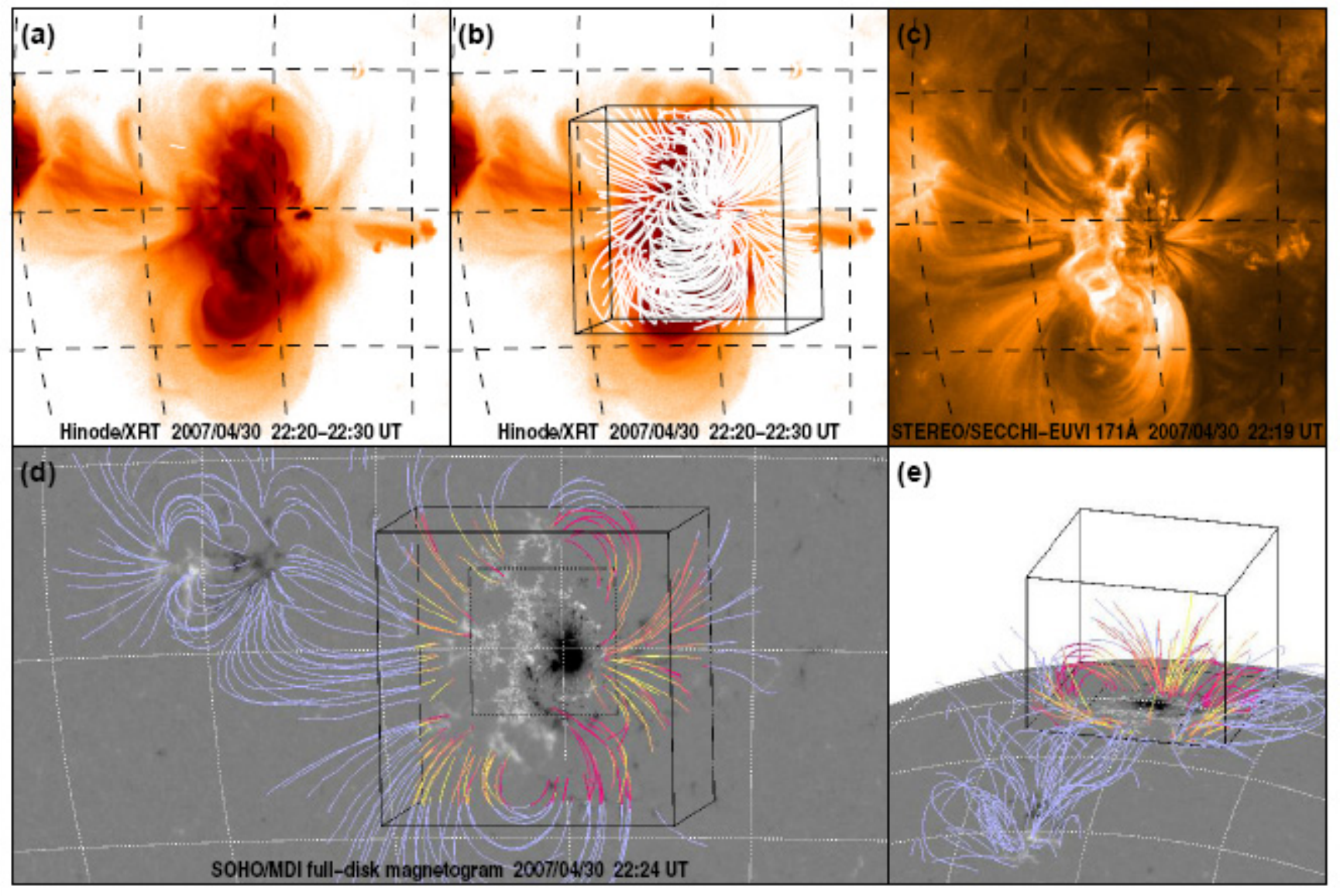}}
\caption{A series of coaligned images of AR~10953. Blue lines are
stereoscopically-reconstructed loops \citep{aschwanden:etal08a}.
Red lines are extrapolated nonlinear force-free field
lines from Hinode/SOT with MDI data outside the Hinode
FOV (the dotted line). The original figure was published
as Figure~1 in \cite{derosa:etal09}.}
\label{nlfff4f1}
\end{figure}}

When presented with complete and consistent boundary conditions,
NLFFF algorithms generally succeed in reproducing the test fields.
However, for a well-observed dataset (a Hinode/SOT-SP vector-magnetogram
embedded in MDI data) the NLFFF algorithms did not yield consistent solutions.
From this study we conclude that one should not rely on a model-field geometry
or energy estimates unless they match coronal observations.
It was concluded that successful application to real solar
data likely requires at least:

\begin{enumerate}
\item Large model volumes with high resolution that accommodate most of
the field-line connectivity within a region and to its surroundings.
\item Accommodation of measurement uncertainties (in particular in
the transverse field component) in the lower boundary condition.
\item `Preprocessing' of the lower-boundary vector field that
approximates the physics of the photosphere-to-chromosphere
interface as it transforms the observed, forced, photospheric
field to a realistic approximation of the chromospheric, nearly-force-free, field.
\item
The extrapolated coronal magnetic field lines should be compared and
verified by coronal observations.
\end{enumerate}

In the meantime some work has been done in reply to these conclusions.
New implementations of the Grad--Rubin and optimization methods do
accommodate the measurement errors; see Sections~\ref{grad_rubin}
and \ref{optimization} for an overview and \cite{wheatland:etal09},
\cite{wiegelmann:etal10a}, and \cite{amari:etal10} for the corresponding
original publications. On the instrumentation side SDO/HMI provides us
with full-disk measurements of the photospheric magnetic field vector,
which should allow us to find suitable large model volumes. The first
vector magnetograms from SDO/HMI have been released at the end of 2011
and currently research on using them for force-free extrapolations is ongoing.

\subsection{Application of nonlinear force-free codes}
\label{nlfff_applications} Despite the difficulties outlined in
Section~\ref{nlfff_consortium} NLFFF-codes have been used to study
active regions in various situations. Several studies deal with the
energy content of the coronal magnetic field in active regions.
\cite{bleybel:etal02} studied the energy budget of AR~7912 before
and after a flare on 1995 October 14 with a Grad--Rubin method and
found that the magnetic energy decreased during the flare. The
magnetic field lines computed from the nonlinear force-free model
seem to be in reasonable agreement with a soft X-ray image from
Yohkoh, as shown in the top panel in Figure~\ref{bleybel}. At least
the nonlinear force-free model seems to agree better with the X-ray
image than a linear force-free and a potential field model shown in
the center and bottom panel, respectively. \cite{regnier:etal02},
also using the Grad--Rubin approach, studied the non-flaring active
region AR~8151 in February 1998 and found that the available free
magnetic energy was not high enough to power a flare. These results
are consistent which the observation in the sense that nonlinear
force-free field lines reasonably agree with coronal observations
and a consistent flaring activity: The particular active regions
flared (not flared) when the free magnetic energy computed with
NLFFF-codes was high enough (too low).
A decreasing free magnetic
energy during flares has been confirmed in several studies.
\cite{thalmann:etal08} and \cite{thalmann:etal08a},
using the optimization approach, found that the force-free energy
before a small C-class flare (observed in active region 10960 on
2007 June 7) was 5\% higher than the potential field energy.
Before a large M-class flare (observed in active region NOAA~10540 in
January 2004) the force-free energy exceeded the potential field
energy by 60\%. In a statistic study, based on 75 samples
extrapolate with the optimization approach, \cite{jing:etal10} found
a positive correlation between free magnetic energy and the X-ray
flaring index. It seems that we can trust that there is a relation
between computed free energy and flaring activity, whereas the
results of Section~\ref{nlfff_consortium}) indicate that one might
not fully trust in the exact numbers of magnetic energies computed
with one NLFFF-code only. Recently \cite{gilchrist:etal12} pointed
out that uncertainties in the vector magnetograms likely result in
underestimating the computed magnetic energy.
NLFFF-codes are, however, a strong tool to
guide the investigation of coronal features. \cite{regnier:etal04},
\cite{valori:etal12}, and \cite{sun:etal12} applied the Grad--Rubin,
MHD-relaxation and optimization approach, respectively and found at
least  qualitatively a good agreement of NLFFF-models with observed
sigmoid or serpentine structures.

\epubtkImage{bleybel}{%
\begin{figure}[htbp]
\centerline{\includegraphics[scale=0.75]{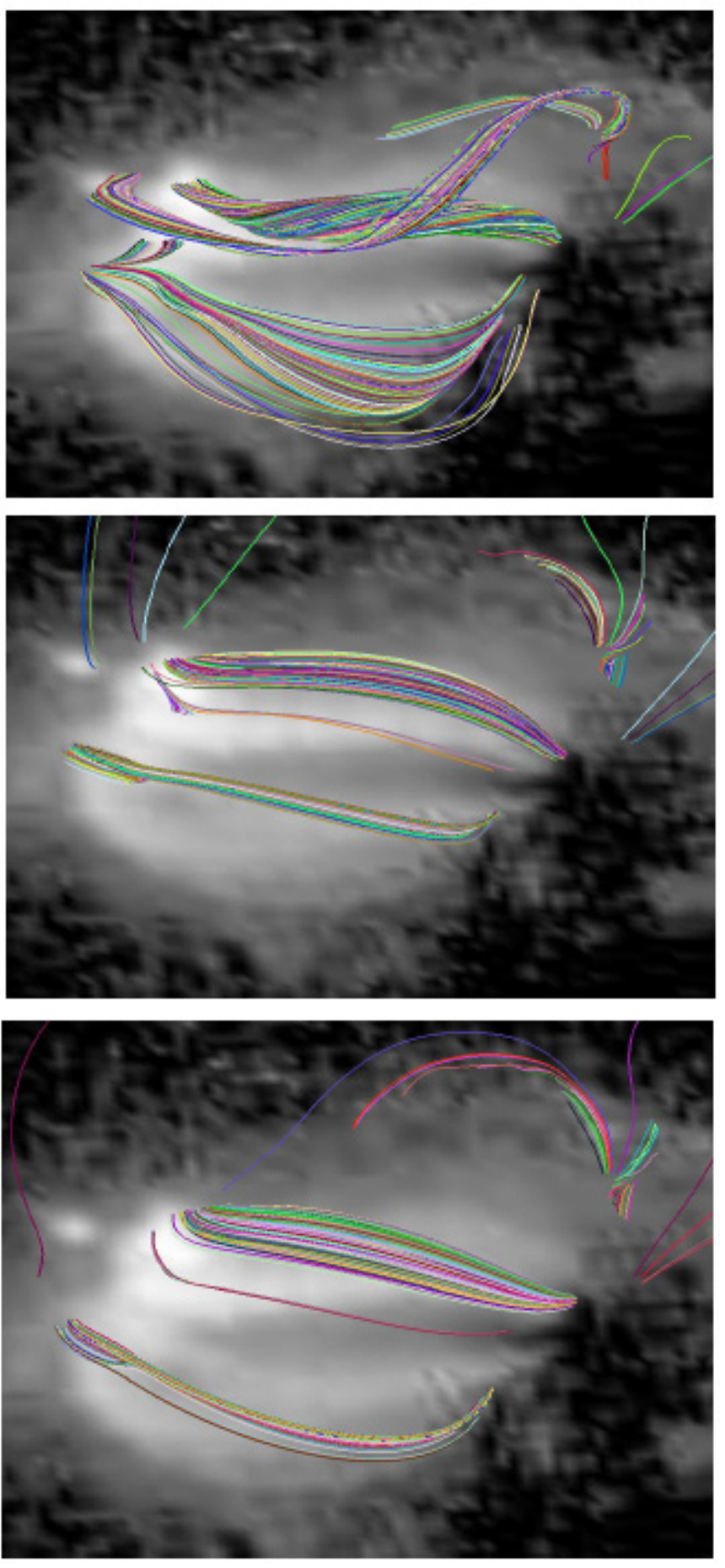}}
\caption{Yohkoh soft X-ray image overlaid with magnetic field lines from
different models: (top) nonlinear force-free, (center) linear
force-free, and (bottom) potential fields. The original figure
was published as Figure~8 in \cite{bleybel:etal02}.}
\label{bleybel}
\end{figure}}

%
\newpage
\section{Summary and Discussion}

In this review we tried to give an overview of force-free magnetic fields,
particularly model assumptions, which are important for understanding
the physics of the solar corona. While the underlying mathematical equations
describe stationary states and look relatively simple, solving them is by no
means a trivial problem because of the nonlinear nature of the problem. Exact
solutions are only available for further simplifications, like linearizing
the equations or to restrict to 1D/2D for the nonlinear case. For force-free
configurations in 3D, we know that (for given flux distributions in the
photosphere) the magnetic field energy is bounded from below by a potential
field. An upper-limit for the energy is more difficult to obtain. While
the Aly--Sturrock conjecture (Section~\ref{aly-sturrock}) claims that the
upper limit is for the configurations with all magnetic field lines open,
\cite{choe:etal02} constructed solutions with energies above the Aly--Sturrock
limit. These configurations contain discontinuities and the debate of the
validity of the Aly--Sturrock limit is ongoing \citep{hu04,wolfson:etal12}.

For practical computations of the 3D-field in the solar corona, one has
to use numerical computations and several codes have been developed,
compared, and applied. As input these codes require measurements of the
magnetic field vector in the solar photosphere. However, the transverse
field component contains an ambiguity in the azimuth, which has to be
resolved before the data can be used for coronal magnetic field modeling.
The accuracy of photospheric measurements is lower for the transverse field
component compared with the line-of-sight field, and in weak field regions
measurements and azimuth ambiguity removal are less trustworthy. Consequently
the majority of coronal force-free field models are carried out in active
regions, although methods for full-disk computations have been developed too.
A further complication of using photospheric measurements as the boundary
condition for force-free extrapolations is that the photospheric magnetic
field is not necessarily consistent with the force-free assumption. Possible
solutions are to use only the vertical magnetic field and the vertical
electric current as boundary conditions, as done for the Grad--Rubin approach,
to preprocess the photospheric measurements with the aim to make them
compatible with force-free and other physical requirements, or to allow
changes of the transverse magnetic field during the iteration of a force-free
field. The latter approach has been implemented in the optimization approach
and allows us to take measurement errors into account.

A major source for future research on force-free fields is SDO/HMI, which
measures the photospheric magnetic field vector on the full disk, which
in principle allows us to compute global coronal models as well as
selecting appropriate isolated active regions with a sufficiently
large field-of-view. Research on Stokes inversion, azimuth ambiguity
removal, and force-free modeling for SDO/HMI data is ongoing. Another
important aspect on coronal modeling is the comparison of force-free
models as extrapolated from photospheric measurements with coronal
images as observed, for example, with the Atmospheric Imaging Assembly
\citep[AIA;][]{lemen:etal12} on SDO. On the one hand, such a comparison
is important to validate the models \citep[see][for details]{derosa:etal09},
and, on the other hand, the 3D models help to interpret the observations.
With the 3D structure of magnetic loops from the models in hand, one
has important tools for modeling of plasma loops, and may gain
understanding of coronal heating and plasma flows along the loops.
Further steps on the research of eruptive phenomena like flares and
CMEs are planned with time-dependent MHD simulations. Force-free models
are planned to be used as initial equilibria, which are disturbed by
photospheric plasma flows (which can be deduced, e.g., from measurements
with SDO/HMI). The temporal evolution and the potential occurrence of
eruptions can be investigated with ideal or resistive MHD simulations
in comparison with observations. Questions are if or to which extent
the configurations remain approximately force-free during eruptions,
the role of thin current sheets and discontinuities, and the energy
and helicity content. We aim to report about the progress in these
aspects in an update of this review in due time.

\section{Acknowledgements}
\label{section:acknowledgements}
TW was supported by DLR-grants 50 OC 0501 and 50 OC 0904.

\newpage


%

\end{document}